\documentclass[fleqn,usenatbib]{mnras}

\usepackage{newtxtext}

\usepackage[T1]{fontenc}

\DeclareRobustCommand{\VAN}[3]{#2}
\let\VANthebibliography\thebibliography
\def\thebibliography{\DeclareRobustCommand{\VAN}[3]{##3}\VANthebibliography}

\usepackage{graphicx}	
\usepackage{amsmath}	
\usepackage{amssymb}	

\usepackage{newtxmath}

\usepackage{siunitx}   
\usepackage{physics}   

\usepackage{booktabs}  
\usepackage{subcaption}  

\newcommand{\mr}{\mathrm}

\newcommand{\DM}{\ensuremath{\mathrm{DM}}}
\newcommand{\tl}{\ensuremath{t_\mr{l}}}

\DeclareSIUnit \pc {pc}
\DeclareSIUnit \erg {erg}
\DeclareSIUnit \dmu {\per\cm\cubed\pc}

\graphicspath{{./}{figures/}}

\title[Mock fast radio burst follow-up]{How limiting is optical follow-up for fast radio burst applications? Forecasts for radio and optical surveys}

\author[J. N. Jahns et al.]{
	Joscha N. Jahns,$^{1}$\thanks{E-mail: \href{mailto:jjahns@mpifr-bonn.mpg.de}{jjahns@mpifr-bonn.mpg.de}}
	Laura G. Spitler,$^{1}$
	Charles R. H. Walker,$^{1}$
	Carlton M. Baugh$^{2}$
	\\
	$^{1}$Max-Planck-Institut für Radioastronomie, Auf dem Hügel 69, D-53121 Bonn, Germany\\
	$^{2}$Institute for Computational Cosmology, Department of Physics, Science Laboratories, Durham University,
	South Road, Durham, DH1 3LE, UK\\
}

\date{Accepted XXX. Received YYY; in original form ZZZ}

\pubyear{2021}

\begin{document}
	\label{firstpage}
	\pagerange{\pageref{firstpage}--\pageref{lastpage}}
	\maketitle
	
	\begin{abstract}
		Fast radio bursts (FRBs) are the first cosmological radio sources that vary on millisecond timescales, which makes them a unique probe of the Universe.
		Many proposed applications of FRBs require associated redshifts.
		These can only be obtained by localizing FRBs to their host galaxies and subsequently measuring their redshifts.
		Upcoming FRB surveys will provide arcsecond localization for many FRBs, not all of which can be followed up with dedicated optical observations.
		We aim to estimate the fraction of FRB hosts that will be catalogued with redshifts by existing and future optical surveys.
		We use the population synthesis code \textsc{frbpoppy} to simulate several FRB surveys, and the semi-analytical galaxy formation code \textsc{galform} to simulate their host galaxies.
		We obtain redshift distributions for the simulated FRBs and the fraction with host galaxies in a survey.
		Depending on whether FRBs follow the cosmic star formation rate or stellar mass, \numrange{20}{40} per cent of CHIME FRB hosts will be observed in an SDSS-like survey, all at $z<0.5$.
		The deeper DELVE survey will detect \numrange{63}{85} per cent of ASKAP FRBs found in its coherent search mode.
		CHIME FRBs will reach $z\sim 3$, SKA1-Mid FRBs $z\sim 5$, but ground based follow-up is limited to $z\lesssim 1.5$.
		We discuss consequences for several FRB applications.
		If $\num{\sim1}/2$ of ASKAP FRBs have measured redshifts, 1000 detected FRBs can be used to constrain $\Omega_\mr b h_{70}$ to within \num{\sim10} per cent at 95 per cent credibility.
		We provide strategies for optimized follow-up, when building on data from existing surveys.
		Data and codes are made available.
	\end{abstract}
	
	
	
	
	\section{Introduction}
	
	Fast Radio Bursts (FRBs) are flashes of radio light coming from distant galaxies. They are a relatively new class of transients \citep{Lorimer2007} that have so far been observed at frequencies between \SI{110}{\MHz} \citep{Pleunis2021a} and \SI{8}{\GHz} \citep{Gajjar2018}.
	Currently, 4 per cent of FRBs have been observed to emit more than once and are therefore classified as repeaters \citep{Spitler2016, CHIME2019}. 
	The larger sample of \citet{CHIME2021} at \SI{600}{\MHz} suggest that repeaters and non-repeaters do indeed have different statistical properties \citep{Pleunis2021}.
	Nevertheless, the source and emission mechanisms are still puzzling \citep[see][for recent reviews]{Petroff2022,Lyubarsky2021}, although the recent detections of an FRB-like burst from the galactic source SGR~1935+2154 \citep{Bochenek2020, CHIME2020, Dong2022} support a connection to magnetars.
	
	Even if their origins remain unclear, FRBs can be used as astrophysical tools in numerous ways. 
	For many applications, the most important quantity that can be measured is the dispersion measure (DM). 
	It is caused by all the free (non-relativistic) electrons along the path between source and observer and manifests as a frequency dependent dispersive delay ($\Delta t\propto \DM\, \nu^{-2}$).
	As a result, FRBs have been proposed as tools for finding the `missing' baryons \citep{McQuinn2014,Prochaska2019,Walters2019}; locating the baryonic matter in the intergalactic medium (IGM), around galaxies, and specifically the Milky Way \citep{Keating2020,Platts2020}; measuring cosmological parameters \citep{Zhou2014,Gao2014}; observing the reionization epochs of H and He \textsc{ii} \citep{Deng2014,Zheng2014,Bhattacharya2021}; measuring intergalactic magnetic fields \citep{Akahori2016,Vazza2018,Hackstein2019}; constraining the abundance of massive compact halo objects \citep{Zheng2014,Munoz2016,Kader2022,Leung2022}; testing Einstein's equivalence principle \citep{Wei2015,Nusser2016,Sen2022}; constraining the photon mass \citep{Wu2016,Bonetti2016}; 
	and others, in particular various applications in the case of strongly lensed (repeating) FRBs \citep{Li2018,Zitrin2018,Wagner2019,Wucknitz2021}.
	
	Many of these applications require or benefit from knowledge of the FRBs' redshifts. 
	For example, the baryons in the IGM are detected via their contribution to the DM ($\DM_\mr{IGM}$) \citep{Ginzburg1973}. 
	On average, it increases with distance, which means that the redshift $z$ is needed as a second distance estimate to determine the baryon density \citep{McQuinn2014}.
	Likewise, a hypothetical photon mass produces a delay that increases with light-travel-time and therefore redshift \citep[see e.g.][]{WeiWu2020}. 
	Cosmological parameters influence the shape of $\langle \DM_\mr{IGM} \rangle(z)$, again requiring $z$ to be measured, although the large DM scatter makes it difficult for this application to compete with other cosmological probes \citep{Walters2018,Jaroszynski2019}.
	The epoch of H reionization is expected to cause $\langle \DM_\mr{IGM} \rangle(z)$ to plateau around $z\sim6$. 
	The real redshift location can most directly be found through the DM and redshift of high-$z$ FRBs \citep[e.g.][]{Beniamini2021}. 
	In addition to these direct applications, localized FRBs also help to learn more about their local environments, and thus, their potential progenitors \citep[][]{Heintz2020,Bhandari2022}.
	In summary, localizations and redshift measurements of FRBs are crucial for unpacking the full potential of FRBs.
	
	To \textit{localize} an FRB, its location needs to be known with arcsecond precision \citep{Eftekhari2017}.
	Only then can the host galaxy be identified in optical or infrared images to below percent level chance coincidence.
	Once the host galaxy is known, its redshift can be measured using spectroscopy.
	A localization via interferometric follow-up observations is possible for FRBs that repeat frequently \citep[e.g.][]{Chatterjee2017,Marcote2020}.
	Most FRBs, however, have not yet been seen to repeat.
	These can only be localized upon discovery, and only if the discovering instrument is an interferometer (and if the FRB has sufficient signal-to-noise).
	Current instruments that localize FRBs on a regular basis are the Australian SKA Pathfinder (ASKAP), Deep Synoptic Array-110 (DSA-110), and MeerKAT.
	
	Upcoming surveys will -- possibly as soon as 2023 -- yield more than a 100 localizations per year.
	At the time of submission, there are only 24 localized FRBs \citep[see e.g.][]{Bhandari2022}.
	These localizations were obtained over the last 3 years, and are dominated by ASKAP.
	However, this number will grow rapidly in the near future as several instrumental updates are currently carried out.
	DSA-110 \citep{Kocz2019} is currently under commission and will begin full operations in the end of 2022 \citep{Ravi2022}.
	ASKAP's CRAFT coherent upgrade \citep[CRACO;][]{James2022b} is being carried out, which will allow searching in the image plane to yield a boost to $\num{\sim100}$ FRBs per year from ASKAP alone.
	It is expected to be operational early 2023.
	The Canadian Hydrogen Intensity Mapping Experiment (CHIME) outriggers are under construction and will provide very-long-baseline interferometry localization of nearly all the \num{\sim500} FRBs per year that CHIME detects \citep{CHIME2021}.
	These outriggers will likely become operational within 2023 \citep{Vanderlinde2019}.
	On timescales of a few years, additional instruments will be built that are capable of localizing similar numbers of FRBs upon discovery.
	Among these are HIRAX \citep{Crichton2022}, GReX \citep{Connor2021}, BURSTT \citep{Lin2022}, CHORD \citep{Vanderlinde2019}, DSA-2000 \citep{Hallinan2019}, PUMA \citep{CVC2019}, the square kilometre array \citep[SKA][]{Dewdney2009}, and ngVLA \citep{Law2018}.
	
	With this many FRBs with arcsecond positions available, the most likely bottleneck to comprehensive cosmological analyses will be optical follow-up observations that provide host galaxy identification and redshift measurements.
	It will be impossible to dedicate the same amount of observing time for each FRB as is allocated currently \citep[][e.g., together invested \SI{4.4}{\hour} of optical follow-up on one FRB]{Simha2020,Chittidi2021}.
	The available time and the follow-up strategy will influence the number of FRBs with known redshift and their redshift distribution.
	Taking the effect of limited observing time into account in a forecast is difficult, as the available telescope time is unknown.
	Therefore previous forecasts of FRB applications have only considered a localized FRB population with simplified redshift distributions.
	These included FRBs at a fixed redshift, following cosmic distributions like the star formation history, or observed distributions of other sources like supernovae or gamma-ray bursts, and recently the simulation of a realistic distribution for ASKAP/CRACO \citep{James2022b}.
	In this work we want to, for the first time, consider the effects of limited optical follow-up.
	Thus we estimate the fraction of future FRBs whose host galaxies will already be contained in optical catalogues, and conversely, the fraction that will need dedicated follow-up observations with optical telescopes.
	
	We first describe the simulations and parameters used to create our synthetic FRB population in Section~\ref{sec:methods}.
	In Section~\ref{sec:results}, we present the resulting redshift distributions for our simulated FRBs, comparing different underlying radio surveys and simulating the effects of redshift distributions on FRB constraints of the cosmic baryon budget.
	In Section~\ref{sec:follow-up} we develop an optimized follow-up strategy, before we discuss limitations of and prospects for our approach in Section~\ref{sec:discussion}. We conclude in Section~\ref{sec:conclusion}.
	
	\section{Survey simulations}
	\label{sec:methods}
	
	The goal in this section is to generate realistic redshift distributions for future observed FRBs and to compute the fraction of them that have identified host galaxies.
	We do this in two steps, which we summarize here. In the first step, we simulate FRBs using the population synthesis code \textsc{frbpoppy}\footnote{\url{https://github.com/davidgardenier/frbpoppy}}.
	It applies telescope and survey selection criteria to a cosmic FRB population and returns the properties of any observed FRBs.
	In this way, we generate mock catalogues for ASKAP, CHIME, and SKA1-Mid.
	In the second step, we draw a host galaxy for each FRB from a database of simulated galaxies created using the \textsc{galform} semi-analytical galaxy formation code. 
	This database contains magnitudes of galaxies in the passbands for a number of relevant optical surveys.
	We use these magnitudes to ascertain whether the host galaxies could be observed in the following four large surveys: the Sloan Digital Sky Survey (SDSS), the DECam Local Volume Exploration survey (DELVE), the Euclid wide survey, or the Vera C.\ Rubin Observatory's Legacy Survey of Space and Time (LSST).
	We repeat this process for each of our selected radio telescopes and for different cosmic FRB distributions.
	For each radio telescope and distribution, we simulate 1000 observed FRBs.
	Different telescope detection rates could be used to scale the numbers relative to each other to generate realistic detection ratios between telescopes, but this is left for more application specific forecasts.
	In this section, we describe the above codes in more detail and discuss the chosen cosmic probability distributions.
	
	\subsection{Simulation of FRBs with \textsc{frbpoppy}}
	\label{sec:frbpoppy}
	
	\begin{figure*}
		\includegraphics[width=\textwidth]{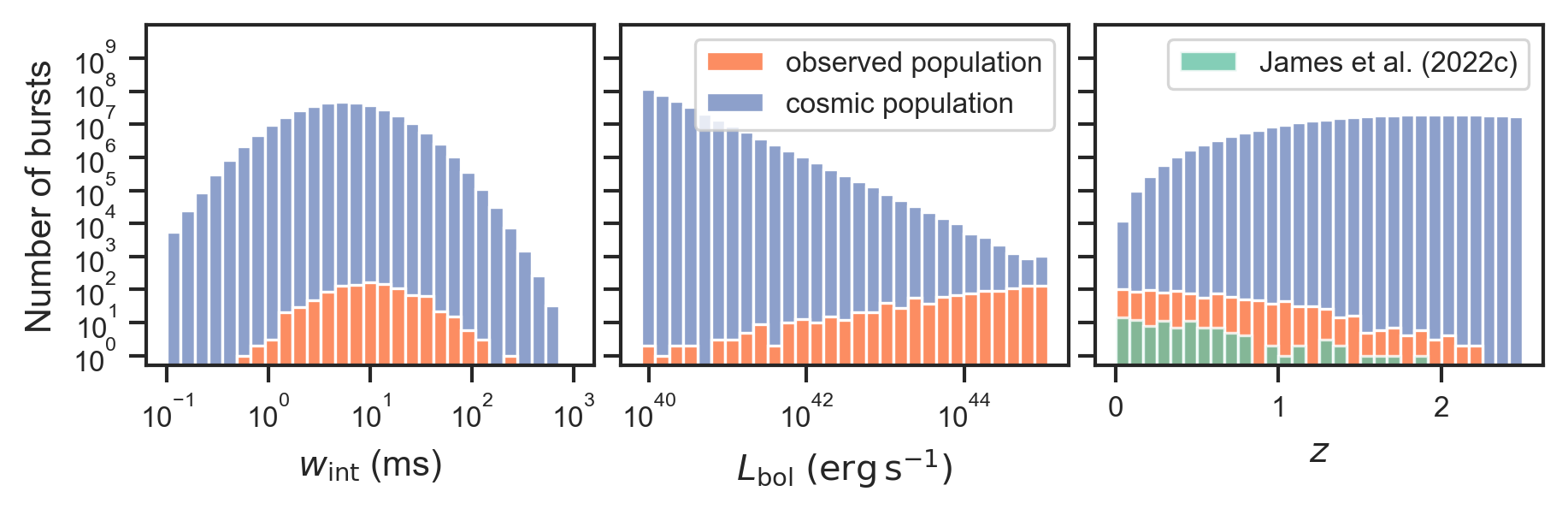}
		\caption{The intrinsic and observed distributions of burst width, luminosity, and redshift that we simulated for ASKAP with the CRACO pipeline. In this example, the FRB population followed the comic star formation rate (SFR). The FRBs simulated in \citet{James2022b} for the same pipeline are shown for comparison. It is apparent in the middle panel how important the maximum luminosity is for the redshift distribution. Even if the luminosity follows a power-law that goes down towards high luminosities, the observed distribution does the opposite, and a large fraction of high luminosity FRBs in the field of view are observed.}
		\label{fig:intrinsics}
	\end{figure*}
	
	The \textsc{frbpoppy} Python package is designed to synthesize FRB populations \citep{Gardenier2019, Gardenier2021}. 
	It is meant to be used to infer the intrinsic FRB properties, but it is also well suited for our forecasts.
	The software synthesizes an FRB population in two steps.
	First, a cosmic population of FRBs is created from intrinsic properties specified by the user, such as the cosmological number density of sources and their luminosity distribution, spectral index, emission range, and pulse widths.
	Second, a telescope and survey is modelled and used to `observe' the FRBs.
	This step requires accurate modelling of telescope parameters including gain, system temperature, beam pattern, and more.
	Below, we describe and justify our choices of parameters.
	For the reader's convenience, values that are used for all surveys are collected in Table~\ref{tab:frbpoppy}, and values that change between our chosen telescopes, or redshift distributions are contained in Table~\ref{tab:params}.
	
	We adopt the cosmological parameters from \citet{Planck2020}.
	For the cosmological FRB number density, we use all three of the models provided by \textsc{frbpoppy}.
	These models are as follows: in the first, the number density follows the redshift evolution of the star formation rate (SFR), in the second, it follows the stellar mass density (SMD), and the third is a toy model, where the number density is constant in the comoving coordinate system.
	Which of the first two models above is correct (or if it is a mix) is still under debate.
	\citet{Zhang2022}, \citet{Qiang2022}, and \citet{Hashimoto2022} use different statistical tests and find that the SMD is favoured.
	However, \citet{James2022} are the only ones that allow for a frequency dependent rate, and they find that the distribution is still consistent with following the SFR.
	For the FRB luminosity function, we use the power-law provided by \textsc{frbpoppy} but the index ($L_\mr{bol,index}$) and range that was found by \citet{James2022}.
	We adopt the log-normal model for the pulse widths from the default population in \citet{Gardenier2019}, with median $\mu_w=\SI{5.49}{\ms}$ and $\sigma_w=2.46$ \citep{James2022a}.
	
	The spectral index is one of the most uncertain properties of FRBs.
	\citet{James2022} and \citet{Shin2022} both infer FRB population parameters under the two interpretations that $\alpha$ is a spectral index, or it expresses how the cosmic rate changes with frequency.
	From the many bursts with limited bandwidth found by \citet{CHIME2021}, it appears that narrowband FRBs dominate the population; thus, we lean towards the rate interpretation.
	We use the index $\alpha=-0.65$ that \citet{James2022} derived under this interpretation from the results of \citet{Macquart2019}.
	This value is also well within the uncertainties of $\alpha$ derived by \citet{Shin2022} for both interpretations ($-1.39\substack{+0.86\\-1.19}$ and $-1.10\substack{+0.67\\-0.99}$ for the spectral index and the rate interpretation, respectively).
	
	\begin{table}
		\caption{Parameters used in \textsc{frbpoppy}.}
		\newcommand{\dittomark}{\textquotedbl\hspace{3em}}
		\label{tab:frbpoppy}
		\begin{tabular}{lrr}
			\toprule
			Parameter          &                    Value & Reference\\ \midrule
			$\nu_\mr{low}$     &           \SI{100}{\MHz} & non-restrictive\\
			$\nu_\mr{high}$    &            \SI{50}{\GHz} & \dittomark\quad\\
			$\alpha$           &                    -0.65 & \citet{James2022}\\
			$L_\mr{bol,max}$   & \SI{3.89e44}{\erg\per\s} & \dittomark\quad\\
			$L_\mr{bol,index}$ &                    -1.05 & \dittomark\quad\\
			$\mu_w$            &           \SI{5.49}{\ms} & \citet{James2022a}\\
			$\sigma_w$         &                     2.46 & \dittomark\quad\\ \bottomrule
		\end{tabular}
	\end{table}
	
	\begin{table*}
		\caption{Parameters used in \textsc{frbpoppy} that differ by survey. $z_\mr{max}$ and $L_\mr{bol,min}$ are chosen as large and as low as possible, respectively, while still having an observable number of FRBs.}
		\label{tab:params}
		\begin{tabular}{llllSS}
			\toprule
			Survey       & Survey model                    & Beam model         & $z$ model   & $z_\mr{max}$ & $L_\mr{bol,min}$ \\ \midrule
			ASKAP/ICS  & \texttt{askap-incoh}            & Gaussian           & SFR         & 1.2          & 2e40             \\
			&                                 &                    & SMD         & 1.2          & 8e39             \\
			&                                 &                    & $V_\mr{C}$ & 1.2          & 5e39             \\
			ASKAP/CRACO  & \texttt{askap-incoh}$^\text{a}$ & Gaussian           & SFR         & 2.5          & 8e39             \\
			&                                 &                    & SMD         & 1.8          & 2e39             \\
			&                                 &                    & $V_\mr{C}$ & 2.0          & 3e39             \\
			CHIME/FRB    & \texttt{chime-frb}              & \texttt{chime-frb} & SFR         & 3.5          & 1e41             \\
			&                                 &                    & SMD         & 2.2          & 1e40             \\
			&                                 &                    & $V_\mr{C}$ & 2.8          & 3e40             \\
			SKA-Mid      & \texttt{ska-mid}                & Gaussian           & SFR         & 6.0          & 3e37             \\
			&                                 &                    & SMD         & 5.0          & 4e36             \\
			&                                 &                    & $V_\mr{C}$ & 6.0          & 3e37             \\ \bottomrule
			\addlinespace[\belowrulesep]
			\multicolumn{6}{l}{\footnotesize{$^\text{a}$ The gain was multiplied by 4.4 and the bandwidth reduced to \SI{288}{\MHz}.}}
		\end{tabular}
	\end{table*}
	
	To calculate the luminosity distribution of our FRB population, we convert the maximum energies inferred by \citet{James2022} and \citet{Shin2022}. The two studies give the maximum energy $E_\mr{max}$ in a \SI{1}{\GHz} band at \SI{1.3}{\GHz} and \SI{600}{\MHz}, respectively. From the data at \SI{1.3}{\GHz}, one can calculate the specific luminosity at frequency $\nu$,
	\begin{equation}
	L_{\nu}=\frac{E_\mr{max}}{\Delta\nu \,\Delta t}\left(\frac{\nu}{\SI{1.3}{\GHz}}\right)^\alpha\,,
	\end{equation}
	where $\Delta\nu=\SI{1}{\GHz}$ is the frequency bandwidth, and $\Delta t$ is a characteristic width of the burst sample.
	For the ASKAP and Parkes FRB sample of \citet{James2022}, we use the median width $\Delta t=\SI{2.67}{\ms}$ reported by \citet{Arcus2021}, and subsequently obtain 
	$
	L_{\SI{1.3}{\GHz}}=10^{35.45^{+0.24}_{-0.48}}\,\si{\erg\per\s\per\Hz}
	$.
	This is in agreement with \citet{Shin2022}, whose result is equivalent to $L_{\SI{1.3}{\GHz}}=10^{35.07^{+0.47}_{-0.46}}\,\si{\erg\per\s\per\Hz}$ (68 per cent confidence limits in both cases).
	Finally, we calculate the bolometric luminosity 
	\begin{equation}
	L_\mr{bol}=\left(\frac{\nu_\mr{high}}{\SI{1.3}{\GHz}}^{1+\alpha} - \frac{\nu_\mr{low}}{\SI{1.3}{\GHz}}^{1+\alpha}\right)L_{\SI{1.3}{\GHz}}\,,
	\end{equation}
	where we use $\nu_\mr{low}=\SI{100}{\MHz}$ and $\nu_\mr{high}=\SI{50}{\GHz}$, to ensure that the emission frequency is not a limiting factor for any of the telescopes.
	
	There are a number of parameters that we do not use here because we simulate a fixed number of FRBs for each survey.
	These include the sky position and absolute rates.
	We simulate the DM separately in Section~\ref{sec:missing}.
	We neglect scattering in this study for two reasons. 
	First, scattering from the host galaxy and Milky Way depends only weakly on the redshift (through a redshift dependent SFR).
	Second, the probability that an FRB will intersect a galaxy is very low \citep{Prochaska2018},
	while the contribution from intervening galaxy haloes to scattering is very uncertain \citep[see e.g.\ the discussion in][]{Ocker2022a}.
	
	We generally use telescope parameters as tabulated in \textsc{frbpoppy} \citep[table 2 of][]{Gardenier2021}, with the exception of ASKAP/CRACO.
	The ASKAP FRB sample will be dominated by the CRACO upgrade as soon as operations begin.
	As it is not yet implemented in \textsc{frbpoppy}, we use the incoherent survey parameters, but multiply the gain by 4.4 and decrease the bandwidth to \SI{288}{\MHz}, as anticipated by \citet{James2022b}.

	\subsection{Host galaxies generated with \textsc{galform}}
	\label{sec:galform}
	
	\textsc{galform} is a semi-analytic model of galaxy formation \citep{Cole2000}.
	The goal of semi-analytic models is to understand the physical processes that govern galaxy formation and evolution.
	The modelling includes 14 different physical processes such as feedback from supernovae and active galactic nuclei \citep{Lacey2016}.
	The gravitational conditions are given by the halo merger tree in which the baryonic physics is implemented.
	This is taken from dark matter only N-body simulations; in the case of the \textsc{galform} version used here, the P-Millennium simulation was used \citep{Baugh2019}.
	The main advantage of semi-analytical models over full hydrodynamical simulations is their speed, which allows the comparison of model galaxies drawn from large numbers of halo merger histories over a wide dynamic range in mass to observed galaxies.
	
	Here we use the \textsc{galform} version from \citep{Lacey2016}, as recalibrated in \citet{Baugh2019}.
	This model includes a detailed treatment of dust absorption, which allows it to produce realistic predictions for the flux from model galaxies in the optical and near infrared.
	Most importantly for us, the optical filters of several telescopes (e.g. SDSS, DECam, Euclid) are applied to generate the model galaxy magnitudes in different bands.
	These magnitudes include the effects of extinction in the host galaxy and are in the observer's reference frame, i.e. they consider the redshifting of the spectrum relative to the filter.
	\textsc{galform} tracks quiescent star formation in galactic disks and bursts of star formation triggered by merger or the motion of gas in dynamical unstable disks.
	In the model used here, bursts are assumed to take place with a top-heavy stellar initial mass function (IMF), whereas a solar neighbourhood IMF is adopted for star formation in disks.
	The model tracks the star formation and mass assembly in a disk and bulge component for each galaxy. Different bulge-to-disk ratios can be associated with different morphological types.
	Apart from the magnitudes, we only need the stellar mass and SFR to randomly draw host galaxies from the population, consistent with their cosmic number density.
	
	Twelve snapshots from the simulation were used between redshifts 0 and 6.011.
	The P-Millennium is a $540/h\,\si{\mega\pc}$ box.
	The model output we used corresponds to a random sampling of merger trees from this volume at the rate of 1/1024.
	The snapshot redshifts and the number of galaxies contained in them are listed in Table~\ref{tab:snapshots}.
	
	\begin{table}
		\caption{\textsc{galform} snapshots that the host galaxies are drawn from.}
		\label{tab:snapshots}
		\begin{tabular}{SSS}
			\toprule
			\text{Snapshot} & \text{Redshift} & \text{Number of galaxies} \\ \midrule
			0               & 0.0             & 182711                    \\
			1               & 0.249           & 192040                    \\
			2               & 0.496           & 201698                    \\
			3               & 0.757           & 212338                    \\
			4               & 1.007           & 221551                    \\
			5               & 1.496           & 238861                    \\
			6               & 2.002           & 254227                    \\
			7               & 2.51            & 251020                    \\
			8               & 3.046           & 233644                    \\
			9               & 3.534           & 212626                    \\
			10              & 4.008           & 190478                    \\
			11              & 6.011           & 95180                     \\ \bottomrule
		\end{tabular}
	\end{table}
	
	We draw a host galaxy for each FRB from the snapshot that is closest to the FRB in redshift space.
	We draw these galaxies weighted either by their stellar mass or SFR, choosing the same that was used for the redshift distribution of FRBs.
	For the redshift distribution following $V_\mr{C}$, we chose the stellar mass.
	
	\subsection{Detections in optical surveys}
	\label{sec:optical}
	
	We wish to assess what fraction of our simulated observed FRBs will have a host galaxy catalogued in one of the surveys SDSS, DELVE, Euclid wide, or LSST. 
	The reason is that this is the fraction of FRBs for which we will get the photometric redshift (photo-$z$) `for free' without needing dedicated follow-up.
	In this analysis, we will concentrate on photo-$z$s.
	This is because the more precise spectroscopic redshifts require much more observation time, resulting in spectroscopic surveys that are usually too shallow or too narrow to cover a significant fraction of FRB hosts.
	Furthermore, a dedicated spectroscopic measurement requires previous detection and identification of the host galaxy.
	Hence, if one requires spectroscopic redshifts for a given method, photometric detection is still the necessary first step.
	
	Photo-$z$s are only an estimate of the true redshift.
	For LSST, the target photo-$z$ precision\footnote{LSST Science Requirements Document available at \url{https://docushare.lsstcorp.org/docushare/dsweb/Get/LPM-17}} is $\sigma_z < 0.02(1 + z)$. 
	The uncertainty is generally redshift dependent \citep[see e.g.][]{Graham2020}, and catastrophic outliers -- substantially inaccurate redshift estimates -- are also possible.
	Such catastrophic outliers could possibly be identified by a mismatch between photo-$z$ and \DM{}, although, care has to be taken to not bias the science that is done with the exact same relation.
	For simplicity, we will assume in this analysis that the uncertainty can be absorbed into other larger uncertainties, like the uncertain host galaxy DM ($\DM_\mr{host}$), and the scatter in $\DM_\mr{IGM}$ that comes from the large scale structure and intervening galaxy haloes.
	In a way, we regard spectroscopic redshifts as a bonus that would improve precision.
	As a motivation, we can compare $\sigma_z$ to the scatter in $\DM_\mr{host}$.
	If we assume $\DM_\mr{host}$ has a log-normal probability with a median of $\DM_0=\SI{100}{\dmu}$ and width parameter $\sigma_\mr{host}=1$, the relative uncertainty of $\sigma_z$ would be 40 per cent of $\DM_\mr{host}$ at $z=1$ and equal to the standard deviation of $\DM_\mr{host}$ around $z=2$. Here we approximated $\langle \DM_\mr{IGM}\rangle\approx\SI{1000}{\dmu}\cdot z$ and used the standard deviation of the log-normal distribution $\DM_0\sqrt{\exp\bigl(2\sigma_\mr{host}^2\bigr)-\exp\bigl(\sigma_\mr{host}^2\bigr)}$.
	
	In order to assess the visibility in optical surveys, the absolute magnitudes $M_{\textsc{G}}=M-5\log(h)$ that \textsc{galform} provides (in the observer frame and including extinction) need to be converted to the apparent magnitudes $m$, as would be observed from Earth.
	This is done via
	\begin{equation}
	m = M_\mr{G} + 5\log(h) - 2.5\log(1+z) + 5\log\left(D_L(z)/\SI{10}{\pc}\right)\,,
	\end{equation}
	where $h$ is the dimensionless Hubble constant and $D_L$ the luminosity distance.
	Note that $M$ is already in the observer frame, and the $-2.5 \log (1+z)$ term is a band shift term from the magnitude definition used in \textsc{galform}.
	The apparent magnitudes are then compared to the survey limits of the numerous bands listed in Table~\ref{tab:overlaps}.
	
	We assume that a redshift can be obtained if a galaxy is visible in all bands.
	We confirmed this simple approach for SDSS by comparing it to the more sophisticated requirements of \citet{Beck2016}. We found that almost no galaxies are excluded by the additional requirements.
	Another reason to refrain from using a specific algorithm to compute photo-$z$s from the simulated magnitudes is the large number of available algorithms that have been developed for LSST \citep[e.g.][]{Schmidt2020}.
	
	\subsection{Survey overlaps}
	\label{sec:overlaps}
	\begin{table*}
		\caption{Optical surveys and their overlaps with FRB surveys.}
		\label{tab:overlaps}
		\begin{tabular}{lllSlrr}
			\toprule
			&		&		&		&		&	\multicolumn{2}{c}{Sky overlap with}			\\
			&		&		&		&		&	ASKAP/SKA	&	CHIME	\\
			Survey	&	Filter	&	Magnitude limits $^\text{f}$	&	{Sky area (\si{\square\deg})}	&	Restrictions $^\text{g}$	&	($\delta < \ang{48}$)	&	($\delta > \ang{-10}$)	\\ \midrule
			SDSS $^\text{a}$	&	u, g, r, i, z	&	22.0, 22.2, 22.2, 21.3, 20.5	&	14555	&		&	$\lesssim\SI{30}{\%}$	&	$\sim\SI{50}{\%}$	\\
			LSST $^\text{b}$	&	u, g, r, i, z, y	&	26.1, 27.4, 27.5, 26.8, 26.1, 24.9	&	18000	&	$\ang{5}>\delta>\ang{-65}$	&	$>\SI{50}{\%}$	&	$\sim\SI{5.5}{\%}$	\\
			Euclid wide survey $^\text{c}$	&	I, Y, J, H	&	26.2, 24.5, 24.5, 24.5	&	15000	&	$|\beta|>\ang{10}$, $|b|>\ang{23}$	&	$>\SI{35}{\%}$	&	$\lesssim\SI{35}{\%}$	\\
			DELVE (DR2) $^\text{d}$	&	g, r, i, z	&	24.3, 23.9, 23.5, 22.8	&	17000	&	$|b|>\ang{10}$, $\delta<\ang{30}$	&	$\sim\SI{50}{\%}$	&	$<\SI{25}{\%}$	\\
			\midrule													
			Pan-STARRS1 survey $^\text{e}$	&	g, r, i, z, y	&	23.3, 23.2, 23.1, 22.3, 21.3	&	31000	&	$\delta > \ang{-30}$	&	$<\SI{70}{\%}$	&	\SI{100}{\%}	\\
			
			\bottomrule
			\addlinespace[\belowrulesep]
			\multicolumn{7}{l}{\footnotesize{$^\text{a}$ \citet{Abazajian2009}, \citet{Alam2015}, \url{https://www.sdss4.org/dr17/scope}}} \\
			\multicolumn{7}{l}{\footnotesize{$^\text{b}$ \citet{Ivezic2019}, \url{https://www.lsst.org/scientists/keynumbers}}} \\
			\multicolumn{7}{l}{\footnotesize{$^\text{c}$ \citet{Euclid2022}, \url{https://sci.esa.int/web/euclid/-/euclid-nisp-instrument}}}\\
			\multicolumn{7}{l}{\footnotesize{$^\text{d}$ DECam Local Volume Exploration survey; \citet{DrlicaWagner2022}}}\\
			\multicolumn{7}{l}{\footnotesize{$^\text{e}$ The Panoramic Survey Telescope and Rapid Response System (Pan-STARRS) is not included in our simulations; \citet{Chambers2016}.}}\\
			\multicolumn{7}{l}{\footnotesize{$^\text{f}$ SDSS: 95 per cent completeness for point sources. LSST: 5$\sigma$ point source depth for stationary sources after 10 years. Euclid: 5$\sigma$ point source depth.}}\\
			\multicolumn{7}{l}{\footnotesize{The DELVE and Pan-STARRS1 survey have inhomogeneous coverage, thus denoted magnitudes are the median and mean 5$\sigma$ point-source depth, respectively.}}\\
			\multicolumn{7}{l}{\footnotesize{$^\text{g}$ $\delta$ denotes the declination, $b$ the Galactic latitude, and $\beta$ the ecliptic latitude.}}
		\end{tabular}
	\end{table*}
	
	We chose the four optical surveys by availability in \textsc{galform} and relevance to FRB surveys.
	SDSS represents a well-established survey with significant legacy data. Situated in the Northern Hemisphere, it is most relevant to CHIME.
	The Pan-STARRS1 survey, which covers almost the entire Northern Sky, was not available in the simulation.
	Its depth is reported as the mean depth, differently from SDSS, and it has one filter that is different, but taking these differences into account, the depth is roughly similar to SDSS.
	The SDSS results are therefore also applicable to Pan-STARRS1, and we refrain from simulating the Pan-STARRS bands additionally to the SDSS bands.
	
	DELVE represents a newer, ongoing survey that is slightly deeper.
	It covers large parts of the Southern Hemisphere and is therefore most relevant to telescopes like ASKAP and MeerKAT.
	With LSST, we consider a wide and deep survey that represents the best that will be available in the foreseeable future.
	As the full survey will only be complete in 10 years (although with yearly data releases), we mainly present LSST with our future radio survey SKA1-Mid.
	
	In the following, we describe how we estimate the overlap between our FRB-searching radio surveys and host galaxy-identifying optical surveys, which we tabulate in Table~\ref{tab:overlaps}.
	Optical surveys observe to equal depths within most of their footprint.
	Therefore, we frame our question as: what fraction of time will our FRB surveys spend within the footprints of our optical surveys?
	
	Most FRB surveys piggyback on other radio surveys.
	These surveys are numerous in the case of ASKAP (and MeerKAT), and only dictate their observing schedules in the near future.
	Our limited knowledge is best described by assuming isotropic coverage of the visible sky for ASKAP and SKA1-Mid.
	Following this assumption, we estimate the FRBs that will be within an optical survey by the fractional overlaps of the visible sky with the optical survey footprints.
	
	The ASKAP telescope is located at a latitude of \ang{-26.7} and can observe sources from declination \ang{-90} to \ang{+48} \citep{Hotan2021}.
	Similarly, MeerKAT (and therefore the future SKA1-Mid), which is situated at latitude \ang{-30.7}, can observe up to declination \ang{+44} \citep{Kapp2016}.
	DELVE is the combination of data from several surveys that were conducted with the Dark Energy Camera (DECam).
	The goal of DELVE is to image the entire Southern Sky, except for the Galactic plane, in four bands, which would eventually yield \SI{\sim26000}{\square\deg} coverage.
	The Vera C.\ Rubin Observatory is located at latitude \ang{-30.2}.
	Its main survey, the LSST, will cover about \SI{18000}{\square\deg} \citep{Marshall2017} from \ang{-65} to about \ang{+5}, excluding the Galactic plane.
	
	The CHIME telescope in the Northern Sky is a transit telescope with a declination dependent beam \citep{CHIME2021,Josephy2021}.
	Since SDSS is also not homogeneous, we do the following estimate.
	We approximate the CHIME detection rate to be constant in declination at $\delta>\ang{0}$ in rough agreement with the results of \citet{Josephy2021}.
	For SDSS, we estimate the coverage from the footprint \citep{Aihara2011} to be $3/4$ at $\delta=\ang{0}$--\ang{40}, $1/2$ at $\delta=\ang{40}$--\ang{70}, and 0 otherwise.
	We estimate about half of CHIME's FRBs to land in the SDSS footprint, yet, we note again that the other half is completely covered by Pan-STARRS1, which is of similar depth.
	
	Euclid's survey area is equally distributed between Northern and Southern Sky, covering $\num{\sim35}$ per cent of the entire sky.
	It is therefore of interest to all FRB surveys.
	However, it targets $z\sim1$ galaxies using one broad optical band (the I band) and three infrared bands (Y, J, H).
	Spectral features that are important for photo-$z$ determination remain in the same band over the full expected redshift range up to $z\sim2$.
	The \SI{4000}{\angstrom} break, for example, will be in the I-band for all galaxies \citep[see e.g.\ section 5.5 of][]{Euclid2022}.
	Euclid will therefore rely on photo-$z$s from optical, ground-based telescopes.
	Keeping this in mind, Euclid can still be interesting for identifying host galaxies as it is the second-deepest survey considered here, after the LSST.
	
	After outlining the survey situation, we want to gauge if the coverage or the depth of optical surveys is the limiting factor.
	Thus, we need to estimate what fraction of the sky is not covered by any optical survey.
	CHIME's visible sky is completely covered by SDSS and Pan-STARRS1, albeit to lower depth in the Galactic plane.
	ASKAP's sky is covered to 50 per cent by the DELVE survey, but SDSS and Pan-STARRS1 cover everything else that is above $\delta=\ang{-30}$.
	This leaves only the Milky Way at $\delta < \ang{-30}$ uncovered, which is about 10 per cent of the total field of view of ASKAP.
	Altogether, the depth of the surveys will be the limiting factor for all radio telescopes.
	
	Throughout the remainder of the paper, we only consider FRBs within the optical survey footprints.
	We leave the absolute number of how many FRBs will be in which optical survey open.
	
	\subsection{Milky Way extinction}
	In the previous section and in this work overall, we do not consider extinction from the Milky Way.
	This simplification is mostly to keep our results independent of the sky direction, except for being either inside or outside of an optical survey.
	This simplification is not always justified \citep{Schlegel1998}\footnote{A tool to estimate extinction is available at \url{https://irsa.ipac.caltech.edu/applications/DUST/}}, in particularly in the galactic plane, where extinction often exceeds \SI{1}{mag};
	for example in the cases of FRB~20180916B \citep{CHIME2019,Marcote2020} and FRB~20210407E \citep{Shannon2023} at galactic latitudes $b=\ang{3.72}$ and \ang{-6.71}, respectively.
	Enhanced scattering of FRBs in the Galactic disk does not significantly affect the FRB detection rates \citep{Josephy2021} and does therefore not reduce the importance of Milky Way extinction.
	However, the Milky Way DM contribution is much higher in the Galactic plane, resulting in much higher $\DM$ uncertainties \citep[see, e.g.]{Price2021}.
	It might therefore be beneficial to exclude FRBs that are in the Galactic plane to avoid potential biases.
	By ignoring Milky Way extinction, we therefore make the hidden assumption that only FRBs outside the Milky Way plane will be used for cosmological applications.
	Considering this, the estimated survey overlaps in Section~\ref{sec:overlaps} are somewhat conservative, because we did not exclude the Galactic plane in the estimates.
	
	\section{Results}
	\label{sec:results}
	
	From the simulations described in Section~\ref{sec:methods} we obtain observed populations of FRBs and their host galaxies.
	The quantities we collect for FRB populations include their redshifts, and host galaxy quantities include their magnitudes in several optical surveys, which informs us which FRBs would have a measured redshift.
	These data provide us with an observed redshift distribution, from which we can directly forecast constraints on cosmological FRB applications.
	We present these results in the following section.
	
	The simulated parameter space of different survey combinations is too large to be fully discussed here, so we limit ourselves to a selection of the results.
	We present the combinations ASKAP/CRACO with DELVE, CHIME with SDSS, and SKA1-Mid with LSST.
	Additional combinations of FRB and optical/infrared surveys, in particular with the Euclid survey, are presented in Appendix~\ref{app:results}.
	
	\subsection{ASKAP}
	
	\begin{figure*}
		\begin{subfigure}{.49\textwidth}
			\includegraphics[width=\columnwidth]{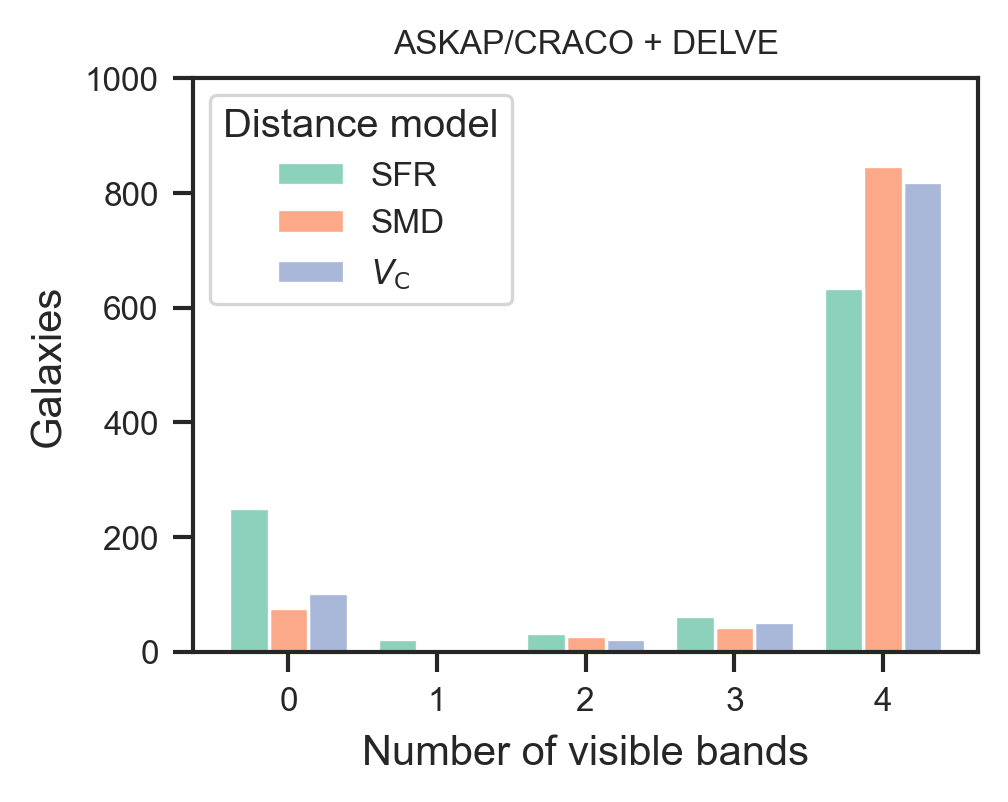}
		\end{subfigure}
		\begin{subfigure}{.49\textwidth}
			\includegraphics[width=\columnwidth]{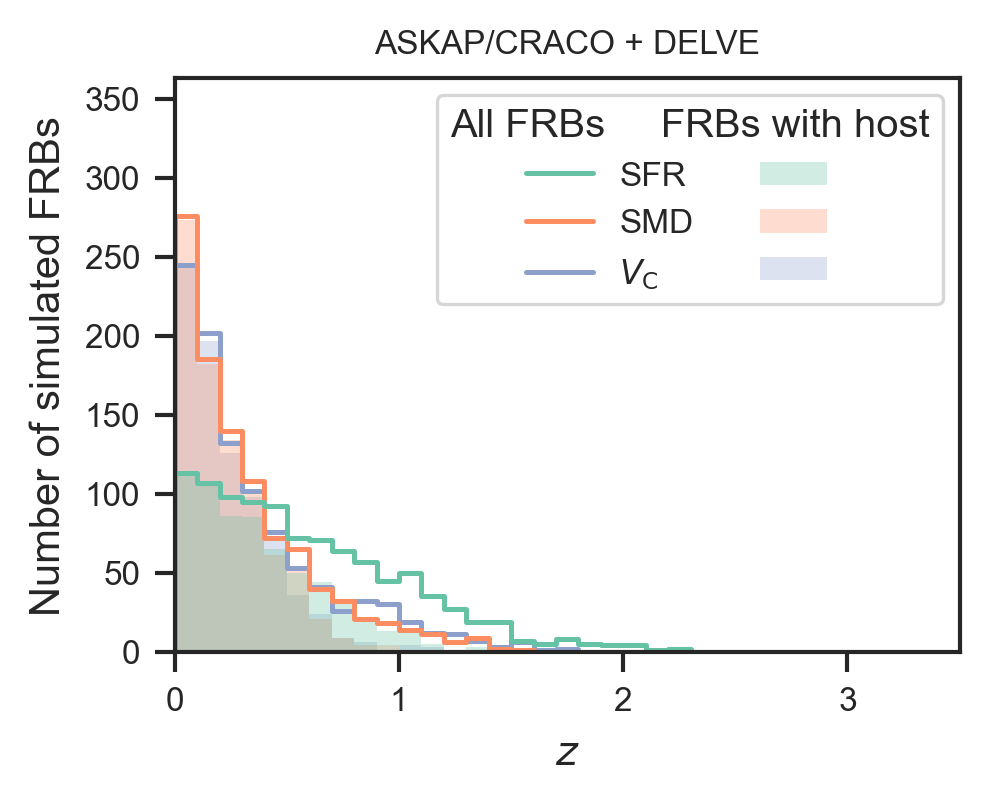}
		\end{subfigure}
		\caption{\textit{Left:} Forecast of the number of DELVE bands in which FRB host galaxies will be visible for 1000 FRBs detected by ASKAP with the CRACO update. Shown are three different models where the intrinsic FRB distance distribution follows the SFR, the stellar mass density (SMD), or the comoving volume ($V_\mr{C}$). DELVE will detect \numrange{\sim63}{85} per cent of ASKAP host galaxies in all bands, depending on the true cosmic FRB distance distribution. \textit{Right:} The redshift distributions of FRBs (lines) and of FRBs whose host was detected in all bands (shaded regions). If FRBs follow the SFR, more are detected at high redshifts, and it will reach up to $z\sim2$. The larger distances result in less detections in all bands compared to other distance models.}
		\label{fig:ASKAP}
	\end{figure*}
	
	The ASKAP telescope is located in the Southern Hemisphere and has a large \SI{30}{\square\deg} field of view thanks to its phased array feeds \citep{Bannister2019}.
	In incoherent sum mode (ICS) its FRB survey (CRAFT) is relatively shallow with all FRBs at $z\lesssim1$.
	The upcoming CRACO mode will be significantly deeper according to our simulations, as is shown in Fig.~\ref{fig:ASKAP}.
	It will detect FRBs up to $z\sim2$ if FRBs follow the SFR, or to $z\sim1.5$ in the two other simulated distance models.
	
	Many host galaxies of ASKAP/CRACO FRBs will be visible in the DELVE survey.
	The numbers that were visible in all bands of the DELVE survey are 634, 847, and 819 out of 1000, for the distance models SFR, SMD, and $V_\mr{C}$, respectively.
	This is also shown in the left panel in Fig.~\ref{fig:ASKAP}.
	Furthermore, only \numrange{7.6}{25} per cent would not be detected in any of the bands, such that the FRBs would have a completely unidentified host.
	
	The FRB population in the right panel of Fig.~\ref{fig:ASKAP} that follow the SFR are clearly distinct from the ones that follow SMD and $V_\mr{C}$.
	The cosmic SFR increases towards its peak around $z\sim2$.
	The effect of this is visible as the FRBs are detected in higher numbers at about $z>0.5$ compared to the populations that follow SMD or $V_\mr{C}$.
	They also reach a higher maximum redshift, but are much less abundant at $z<0.2$.
	The same can also be seen for other surveys (see next sections).
	
	It can be difficult to unambiguously identify higher $z$ FRB hosts because of chance coincidence rates, even when their host galaxy is visible.
	Calculating the chance coincidence, e.g.\ via \textsc{path} \citep{Aggarwal2021a}, requires the probability that a galaxy is visible, prior to consulting optical images at the sky position.
	The distribution in Fig.~\ref{fig:ASKAP} can be interpreted as this prior probability distribution for a given redshift.
	Around $z\sim 0.7$, the probability drops below 0.5 for ASKAP/CRACO.
	This low prior probability can become an issue, in particular for ASKAP FRBs, because ASKAP's localization precision is sometimes on the order of several arcseconds \citep[see e.g.]{Macquart2020}.
	With the resulting high chance coincidence probabilities, secure associations will be difficult for distant FRBs.
	Since it would require ray-tracing simulations, we do not further consider these effects in this study.
	
	\subsection{CHIME}
	
	\begin{figure*}
		\begin{subfigure}{.49\textwidth}
			\includegraphics[width=\columnwidth]{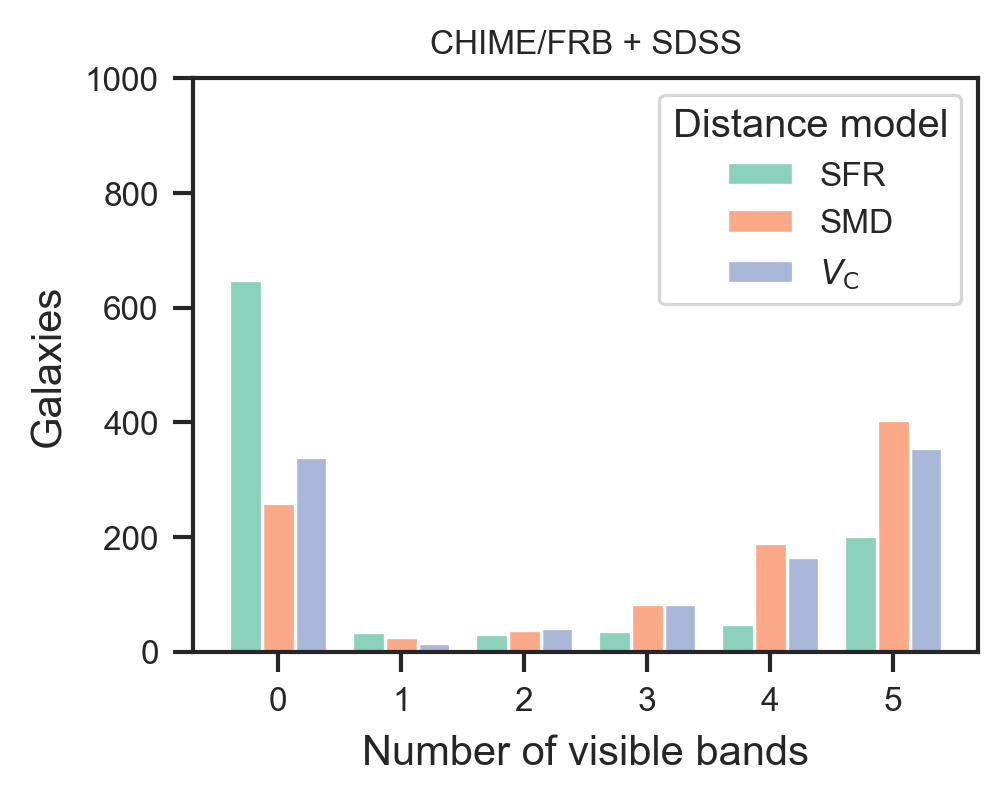}
		\end{subfigure}
		\begin{subfigure}{.49\textwidth}
			\includegraphics[width=\columnwidth]{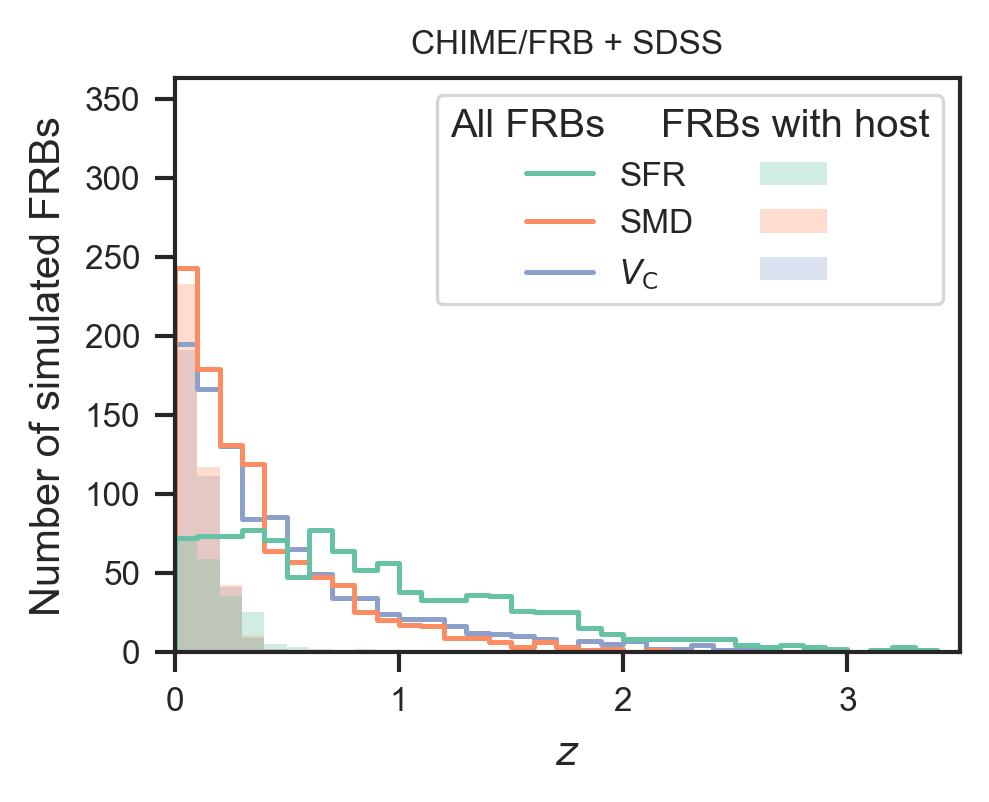}
		\end{subfigure}
		\caption{Like Fig.~\ref{fig:ASKAP}, but for the Northern Sky, with CHIME as FRB instrument and SDSS as the optical survey. SDSS only contains the hosts of FRBs at $z\lesssim0.4$, which only covers \numrange{20}{40} per cent of the CHIME telescopes FRBs, depending on the distance model. As in Fig.~\ref{fig:ASKAP} this does not include the FRB fraction that will be outside the SDSS footprint. Although, this fraction will be covered by Pan-STARRS1, which is of similar depth.
		}
		\label{fig:CHIME}
	\end{figure*}
	
	For CHIME, once it can localize FRBs, the situation will be very different.
	The results of our simulations are shown in Fig.~\ref{fig:CHIME}.
	CHIME is more sensitive than ASKAP and detects FRBs up to $z\sim3$ if FRBs follow SFR, and up to $z\sim2$ if they follow SMD.
	At the same time, the SDSS is shallower than DELVE.
	This results in only 20--40 per cent of CHIME FRBs having their host galaxies detected in all bands of SDSS, while for 26--65 per cent no host can be identified.
	
	Compared to ASKAP, CHIME will detect FRBs to higher redshifts.
	The host galaxies predicted to be seen in SDSS are not only a small fraction of the total, but also all fall below $z=0.5$.
	We will explore the impact of having only low-$z$ FRBs in Section~\ref{sec:missing}.
	The coverage gets worse if we consider that a significant fraction of FRBs will be outside the SDSS footprint.
	Although, the Pan-STARRS1 survey, which we did not simulate, covers the entire CHIME sky and its mean sensitivity lies between the SDSS and DELVE surveys.
	Either way, we can only harvest the signal in CHIME's high-redshift FRBs for cosmological analysis, if we follow them up with dedicated optical observations.
	
	The low prior probability of higher-$z$ FRBs to have a visible host is less problematic for CHIME.
	With its long baseline outrigger stations, it will have a very precise localization precision.
	Yet, some FRBs in the outskirts of their host galaxies will still be difficult to associate with their host.
	A visible galaxy therefore does not guarantee a host identification.
	
	\subsection{SKA1-Mid}
	
	\begin{figure*}
		\begin{subfigure}{.49\textwidth}
			\includegraphics[width=\columnwidth]{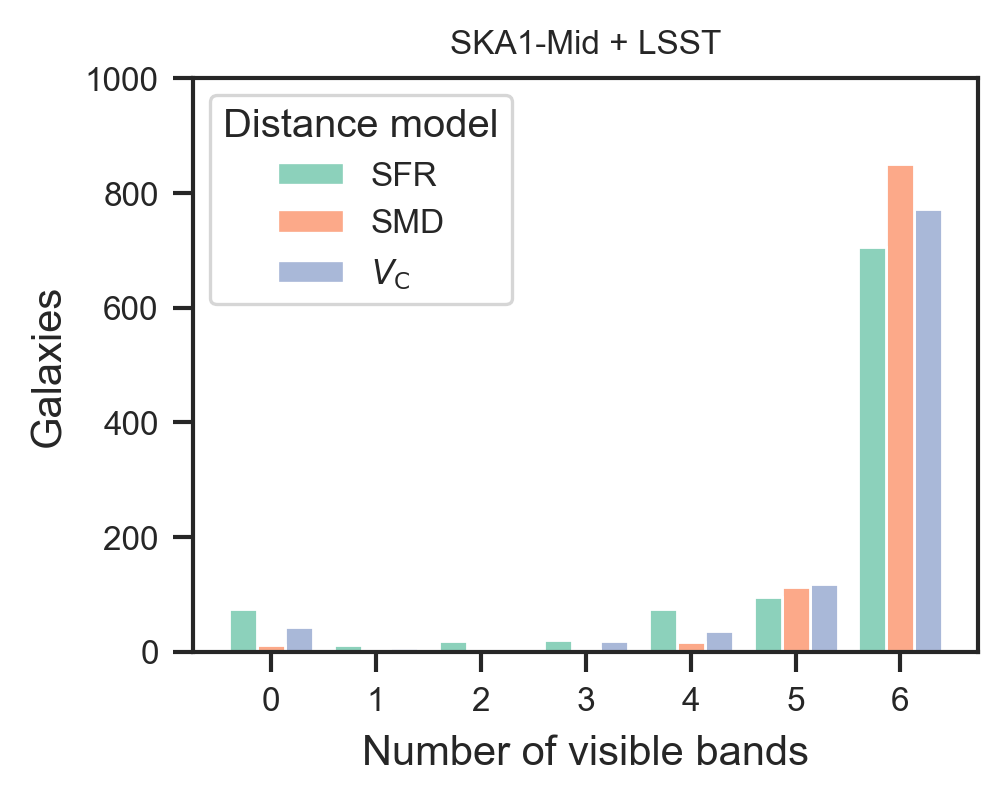}
		\end{subfigure}
		\begin{subfigure}{.49\textwidth}
			\includegraphics[width=\columnwidth]{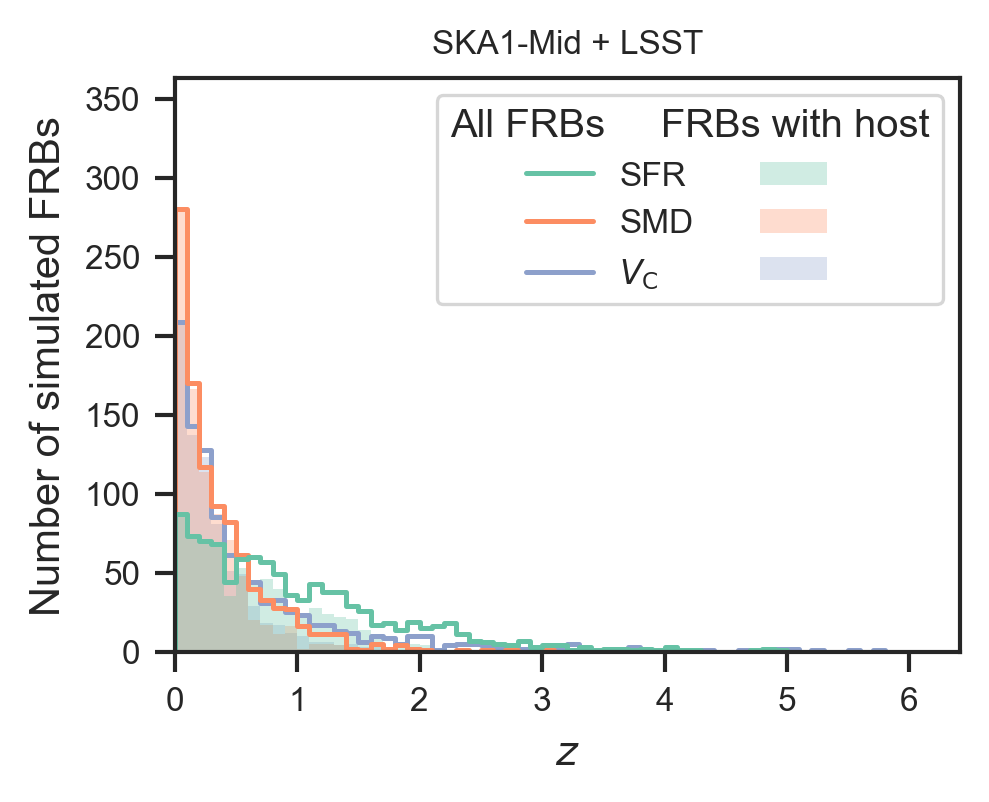}
		\end{subfigure}
		\caption{Similar to Fig.~\ref{fig:ASKAP} for SKA. Despite SKA's sensitivity, still \numrange{71}{85} per cent -- depending on the distance model -- will be detected in all bands of the Vera Rubin Observatory. The SKA barely reaches $z\sim5$ in two models, which is still not sufficient to reach the H reionization epoch at $z\sim6$.}
		\label{fig:SKA}
	\end{figure*}
	
	The results for SKA1-Mid are shown in Fig.~\ref{fig:SKA}.
	SKA1-Mid will be at a similar latitude as ASKAP, but \num{\sim25} times more sensitive \citep{Dewdney2009}.
	The larger FRB distances result in about 71--85 per cent of hosts being visible in all LSST bands in the final data release that will be published after 10 years of observation.
	The visible fraction of host galaxies is decreasing towards $z\sim2$.
	The FRB redshifts observed reach a maximum of $\sim 5$ if FRBs follow SFR or $V_\mr{C}$, but only $z\sim3$ if they follow SMD.
	
	\subsection{Euclid}
	The Euclid results are relevant to all radio surveys, although as we discussed in Section~\ref{sec:overlaps}, Euclid alone cannot obtain photo-$z$s.
	The figures that include Euclid are most interesting in direct comparison to the other optical telescopes, therefore we only present them in Fig.~\ref{app:results}.
	
	As the limiting magnitudes suggest, the results show that Euclid is more sensitive than DELVE.
	Surprisingly, in the cases where FRBs follow the SMD or the $V_\mr{C}$ Euclid also detects a higher number of host galaxies in all bands than LSST.
	An investigation of the visible LSST bands shows that it is almost always the LSST u-band where galaxies are no more visible at higher distances.
	The number of galaxies that are not visible in any band are very similar for Euclid and LSST.
	Another similarity to LSST is that galaxies that are visible in all bands are limited to $z\lesssim 1.5$.
	
	\section{Constraining missing baryons}
	\label{sec:missing}
	
	After simulating different FRB populations, we want to use them as mock observations to forecast constraints on the cosmic baryon density.
	We will do these forecast for the FRBs simulated for ASKAP/CRACO as these will dominate the FRB population in the next 1--2 years, and for CHIME to illustrate the influence of differently distributed FRBs.
	We chose the Bayesian MCMC simulations of \citet{Macquart2020} as the method to constrain the cosmic baryon content.
	For this purpose, we first draw a DM from the same probability distributions that the model of \citet{Macquart2020} assumes.
	In principle, the DM can be split into different contributions, which are difficult to disentangle observationally.
	For this analysis, we express it as the DM from the host galaxy $\DM_\mr{host}$, the intergalactic medium $\DM_\mr{IGM}$, and the Milky Way $\DM_\mr{MW}$, which yields
	\begin{equation}
	\DM = \DM_\mr{MW} + \DM_\mr{IGM} + \frac{\DM_\mr{host}}{1+z}\,.  \label{eq:DM}
	\end{equation}
	In the following, we assumed that contributions from the Milky Way can be sufficiently modelled and therefore only consider DM contributions from the IGM and the host galaxy.
	
	The method of constraining the missing baryons is based on the Macquart relation, which describes the mean DM from the intergalactic medium \citep{Deng2014, Zhou2014}
	\begin{equation}
	\langle\DM_\mr{IGM}\rangle(z) = \frac{3c\,\Omega_\mr{b}H_0f_\mr{IGM}}{8\pi Gm_\mr{p}} \int_0^z \frac{(1+z)[\frac{3}{4}X_\mr{H}(z)+\frac{1}{8}X_\mr{He}(z)]}{\sqrt{\Omega_\mr{m}(1+z)^3+\Omega_\Lambda}}\,\label{eq:Macquart}
	\end{equation}
	where $\Omega_\mr b$ is the cosmic baryon density, $H_0$ the Hubble constant, $f_\mr{IGM}$ the fraction of baryons residing in the IGM, $m_\mr p$ the proton mass, $X_\mr{H}$ and $X_\mr{He}$ the ionization fractions of hydrogen and helium, $\Omega_\mr{m}$ the cosmic matter density, and $\Omega_\Lambda$ the cosmic energy density.
	
	We drew $\DM_\mr{host}$ from a log-normal distribution,
	\begin{align}
	p(\DM_\mr{host}|\DM_0,\sigma_\mr{host}) = &\frac{\log_{10}(e)}{\DM_\mr{host}\sigma_\mr{host}\sqrt{2\pi}} \\
	&\times \exp\left(-\frac{(\log_{10}\DM_\mr{host}-\log_{10}\DM_0)^2}{2\sigma_\mr{host}^2}\right)\,,
	\end{align}
	where we chose a median of $\DM_0=\SI{100}{\dmu}$ and $\sigma_\mr{host}=0.43$, in accordance with the values found by \citet{James2022} and \citet{Shin2022}.
	We drew $\DM_\mr{IGM}$ from
	\begin{equation}
	p_\mr{cosmic}(\Delta)=A\Delta^{-\beta}\exp\left(-\frac{(\Delta^{-\alpha}-C_0)^2}{2\alpha^2\sigma_\mr{DM}^2}\right)\,,\quad \Delta=\frac{\DM_\mr{IGM}}{\langle\DM_\mr{IGM}\rangle}\,,
	\end{equation}
	where we chose $\alpha=3$, $\beta=3$, and $\sigma_\mr{DM} = F/\sqrt{z}$, with $F=0.2$ \citep{Macquart2020}. $A$ and $C_0$ are not free parameters, but determined by the condition $\langle\Delta\rangle=1$ and the normalization.
	Note that in the method of \citet{Macquart2020} the degeneracy between $f_\mr{IGM}$ and $\Omega_\mr bh_{70}$ has been broken, but $\Omega_\mr b$ and $h_{70}$ are still degenerate and the product is measured.
	
	\begin{figure}
		\includegraphics[width=\columnwidth]{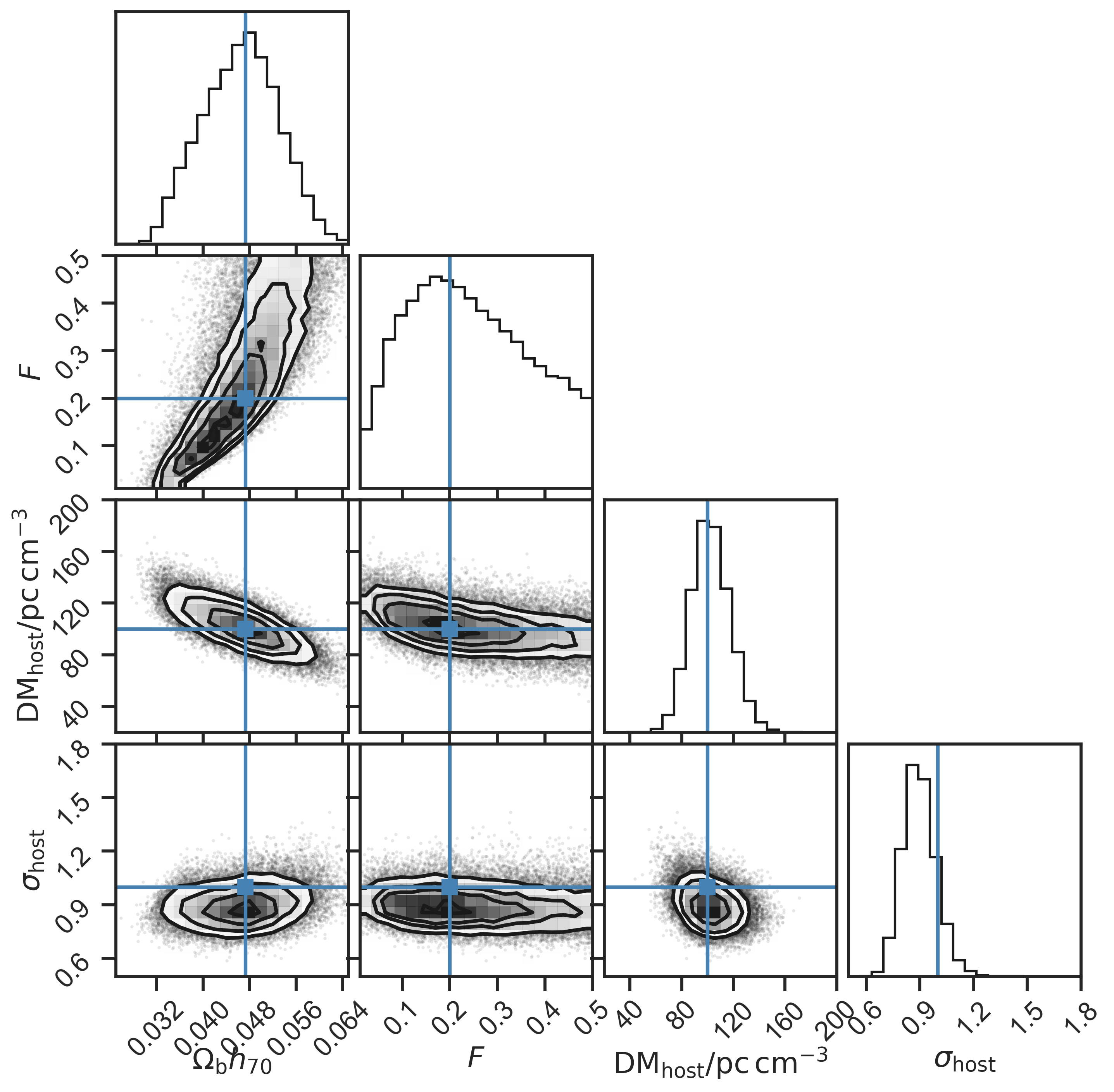}
		\caption{Outcome of an MCMC simulation using the method of \citet{Macquart2020}, for the 124 FRBs that are visible in all bands of SDSS out of \num{1000} simulated CHIME FRBs following the SFR (green shaded region in Fig.~\ref{fig:CHIME}, right panel). Blue lines mark the input values. Contours are at 20, 40, 60, and 80 per cent confidence.}
		\label{fig:corner}
	\end{figure}
	\begin{figure}
		\includegraphics[width=\columnwidth]{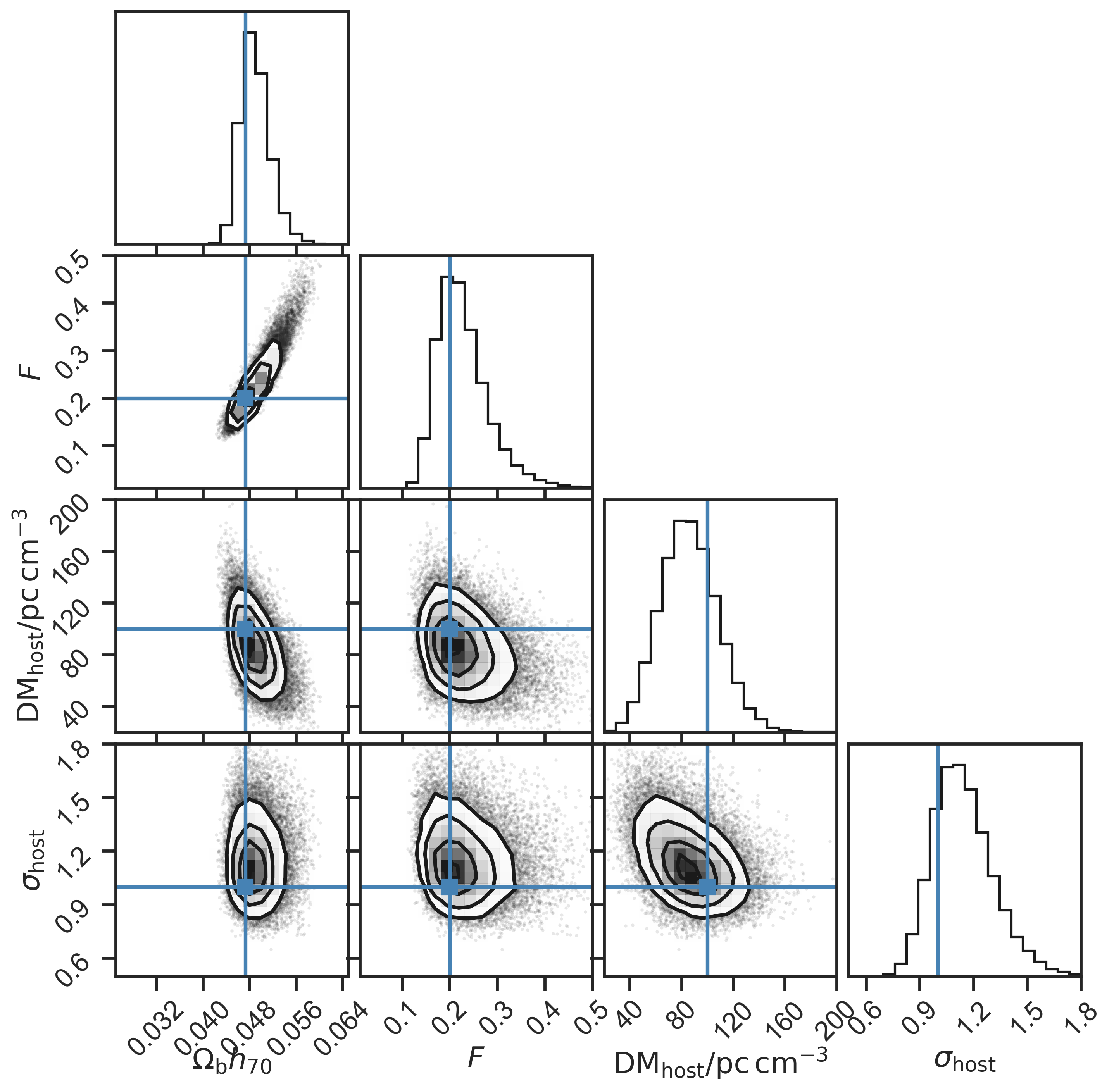}
		\caption{Like Fig.~\ref{fig:corner}, but for comparison 124 FRBs are randomly drawn from the 1000 CHIME FRBs (from the distribution marked by the solid green line in Fig.~\ref{fig:CHIME}, right panel) and assumed to be localized. The cosmological parameters $F$ and $\Omega_\mathrm{b} h_{70}$ are tighter constrained than in Fig.~\ref{fig:corner}, while host galaxy parameters $\DM_\mr{host}$ and $\sigma_\mr{host}$ are less constrained.}
		\label{fig:corner2}
	\end{figure}
	
	\begin{figure}
		\includegraphics[width=\columnwidth]{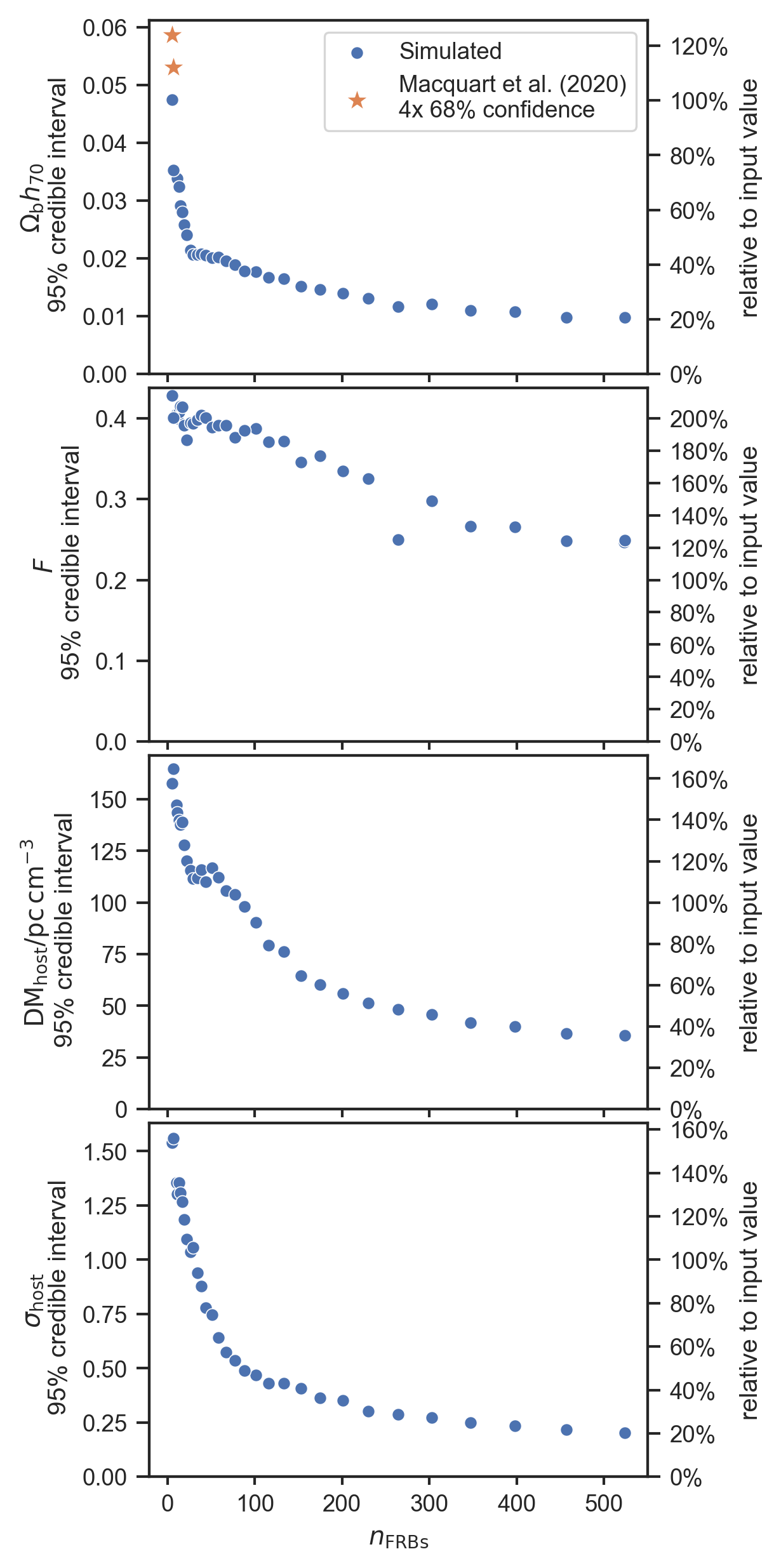}
		\caption{Evolution of the size of the credible interval (specifically the highest density interval) of the four parameters with growing numbers of FRB localizations.
			This forecast is for ASKAP/CRACO with localizations in the DELVE survey.
			For a Gaussian probability function, the 95 per cent credible interval is equivalent to $2\cdot2\sigma$, so we over plot this value as measured from real data.
		}
		\label{fig:Obh}
	\end{figure}
	
	\subsection{Influence of low-$z$ FRBs}
	We do this procedure for the 124 simulated CHIME FRBs that were visible in all SDSS bands.
	To see the influence of the low-$z$ limitation that SDSS imposes on the sample, we repeat the simulations for 124 FRBs that are instead randomly drawn from all the simulated 1000 CHIME FRBs.
	
	The outcomes of the two cases are shown in Fig.~\ref{fig:corner} and Fig.~\ref{fig:corner2}.
	Compared to the results of \citet{Macquart2020} derived from 5 FRBs, the plot shows big improvements in the constraints of all parameters.
	Interestingly, there is a large difference between the two cases.
	Fig.~\ref{fig:corner2} shows much tighter constraints on the cosmological parameters $\Omega_\mr bh_{70}$ and $F$.
	This is the effect of FRBs from higher redshifts carrying a stronger cosmological signal compared to scatter from the inhomogeneous IGM.
	Surprisingly, it is also clear that the low-$z$ FRBs in Fig.~\ref{fig:corner} constrain the host galaxy parameters $\DM_\mr{host}$ and $\sigma_\mr{host}$ better than the population in Fig.~\ref{fig:corner2}.
	This must be a combination of $\DM_\mr{host}$ getting lower with $(1+z)^{-1}$ in Equation~(\ref{eq:DM}), and of less absolute scatter from the IGM at low redshifts.
	The ratios of the 95 per cent credible intervals of the two runs are 2.6, 2.2, 0.7, and 0.5 for $\Omega_\mr bh_{70}$, $F$, $\DM_\mr{host}$, and $\sigma_\mr{host}$, respectively.
	
	\subsection{Evolution of constraints}
	
	We want to see how the constraints on different parameters evolve with the number of FRBs.
	We use the simulated ASKAP/CRACO FRBs that were visible in DELVE, with the distance distribution following the SFR.
	Starting with five FRBs, we consecutively add more FRBs to our detected total up to the maximum of 524 in this run, and we repeat the Bayesian analysis.
	Fig.~\ref{fig:Obh} shows how the size of the 95 per cent credible interval of all four parameters evolves with the number of FRBs.
	The constraints on $F$ only seem to go down linearly, probably due to it still being somewhat degenerate with $\Omega_\mr bh_{70}$.
	The other parameters seem to follow $1/\sqrt{n_\mr{FRBs}}$ laws like quantities with Gaussian distributed uncertainties.
	
	The maximum simulated amount of 1000 ASKAP/CRACO FRBs that resulted in \num{524} hosts in DELVE yields a 95 per cent credible interval of 0.01 for $\Omega_\mr bh_{70}$, which is 21 per cent relative to the input value and roughly equivalent to a 10 percent 2$\sigma$ uncertainty.
	
	\section{Follow-up optimization}
	\label{sec:follow-up}
	
	Dedicated optical follow-up will be needed for galaxies that are either not in survey footprints or too dim.
	Apart from this, spectroscopic follow-up is needed to get precise redshifts of identified hosts and improve uncertainties.
	We investigate in this section how to optimize optical follow-up from theoretical considerations and from our simulations.
	
	\subsection{Theoretical considerations}
	The most important quantity that needs to be considered when seeking to optimize FRB follow-up campaigns is the redshift of the FRBs.
	As different cosmological applications require different FRB redshift populations, sources should be targeted on the basis of these requirements.
	For example, the detection of the epochs of He \textsc{ii} and H reionization require FRBs at $z\gtrsim3$ and 6 respectively.
	For other applications that rely on $\DM_\mr{IGM}$ two effects have to be balanced.
	On the positive side, FRBs that are further away have a higher $\DM_\mr{IGM}$ signal relative to its variance.
	This was first shown by \citet{McQuinn2014}, the average $\DM_\mr{IGM}$ increases faster with redshift than the variance from large scale structure or intervening galaxy haloes \citep[see also][]{Prochaska2019}.
	The variance due to $\DM_\mr{host}$ even gets lower.
	However, this must be considered against the fact that more distant galaxies will, on average, need more observing time.
	
	The observing time that is needed to get a fixed signal-to-noise ratio (SNR) is $t_\mr{obs}\propto S^{-2}$, where $S$ is, in our case, the flux of a host galaxy.
	This flux depends on luminosity $L$ and luminosity distance $D_L$, as $S=L/(4\pi D_L^2)\,.$
	It follows that the expected observation time is
	\begin{equation}
	t_\mr{obs}\propto D_L^4\,,\label{eq:tobs}
	\end{equation}
	for sources whose mean luminosity is constant with z.
	This first order estimate suggests that the increasing observing time dominates the effect of a higher cosmological signal, suggesting it may always be preferable to target close FRBs first.
	
	\subsection{Photometric observing time}
	
	In the following, we will compare this theoretical expectation with our simulations.
	The CHIME/FRB survey with SDSS is the radio/optical combination that requires the most extra follow-up, as many FRBs have no observed host galaxy.
	It is therefore well suited to test different follow-up strategies, and we use it in the following in the version where FRBs follow the SFR.
	To obtain realistic follow-up times, we assume a 10-m optical telescope with two observing systems, a photometer and a slit spectrometer.
	
	We calculate the follow-up time needed for each galaxy for the example photometric and spectroscopic systems following chapter 17.3 of \citet{Schroeder2000}, partly in the notation of \citet{Poggiani2017}. The spectrometer will be considered in Section~\ref{sec:spectro}. Here, we assume that we want to detect each galaxy in a single photometric band of width $\Delta\lambda=\SI{100}{\nm}$, with a target $\mr{SNR}$ of 10. We use the galaxy magnitudes in the simulated SDSS r-band, and the galaxy sizes fixed to about \SI{10}{\kilo\pc}. We assume we are in the background limited regime where the observing time simplifies to
	\begin{equation}
	t_\mr{obs} = \mr{SNR}^2\frac{B}{Q \kappa^2 S^2}\,,\label{eq:snr}
	\end{equation}
	where $S$ is the galaxy flux, $B$ the background flux, $Q$ the quantum efficiency of the detector, and $\kappa$ accounts for losses, not included in the system transmittance. The fluxes are related to the magnitudes by
	\begin{align}
	S &= N_\mr{p}\tau\frac{\pi}{4}(1-\varepsilon^2)D^2\Delta\lambda \cdot 10^{-0.4m}\quad \text{and} \\
	B &= N_\mr{p}\tau\frac{\pi}{4}(1-\varepsilon^2)D^2\Delta\lambda \cdot 10^{-0.4m_B}\phi\phi'\,, \label{eq:background}
	\end{align}
	where $N_\mr{p}=\SI{e4}{photons\per(\s\,\cm\squared\nm)}$ is the magnitude to flux conversion factor, $\tau$ the transmission efficiency, $\varepsilon$ the obscuration factor, $m$ the galaxy magnitude, $m_B$ the sky background magnitude per solid angle, and $\phi\phi'$ the galaxy solid angle. For our example telescope we assume \citep[following][]{Schroeder2000} $Q=0.8$, $\kappa=0.8$, $\tau=0.3$, $\frac{\pi}{4}(1-\varepsilon^2)=0.7$, $D=\SI{10}{\m}$, $m_B=\SI{22}{mag\per\arcsec\squared}$, and $\phi\phi'=\SI{4}{\arcsec\squared}(\SI{1}{\giga\pc}/D_\mr{A})^2$ approximately corresponding to the above-mentioned \SI{10}{\kilo\pc} diameter, with the angular diameter distance $D_\mr{A}$.
	
	\subsection{Spectroscopic observing time}
	\label{sec:spectro}
	
	To calculate the observing time needed for spectroscopy, we take an example split spectrometer.
	The time can be calculated from Equations (\ref{eq:snr}) to (\ref{eq:background}) with two modifications.
	First, the bandwidth $\Delta\lambda$ is now the width of the line of interest, we assume it to be $\Delta\lambda_\mr{line}=\SI{1}{\nm}$.
	Second, the slit might not cover the whole galaxy, in that case the dimensions of the slit and the galaxy's surface brightness will determine $S$.
	We assume that the slit is long enough to cover the whole galaxy, but its width is not.
	The observed flux is then $S_\mr{spec}\approx S\frac{\Delta\lambda_\mr{line}}{\Delta\lambda}\frac{\phi_\mr{slit}}{\phi}$, with $\phi_\mr{slit}$,
	the projected width on the sky, given by $\phi_\mr{slit}=\frac{w'}{rDF_2}$, where $w'$ is the slit, reimaged on the camera focus, $r$ the anamorphic magnification, and $F_2$ the ratio of the camera optics' focal length to the diameter of the collimated beam, incident on the disperser.
	We use again the values from \citet{Schroeder2000}: $w'=\SI{30}{\um}$, $r=0.9$, and $F_2 = 1.5$, which yield $\phi_\mr{slit}=\SI{0.46}{\arcsec}$.
	
	We use the photometric magnitudes, therefore the calculated time is for the continuum and would be less for specific emission lines.
	Additionally, the assumed SNR of 10 is higher than needed for a redshift measurement, which only requires line positions.
	The calculated time is therefore a conservative value.
	The reader may scale the resulting times with a constant factor for differing assumptions.
	Variations in intrinsic galaxy sizes, which we assumed to be fixed, might cause some additional scatter in the required observing times.
	
	\subsection{Results}
	\label{sec:obstimes}
	
	\begin{figure}
		\includegraphics[width=\columnwidth]{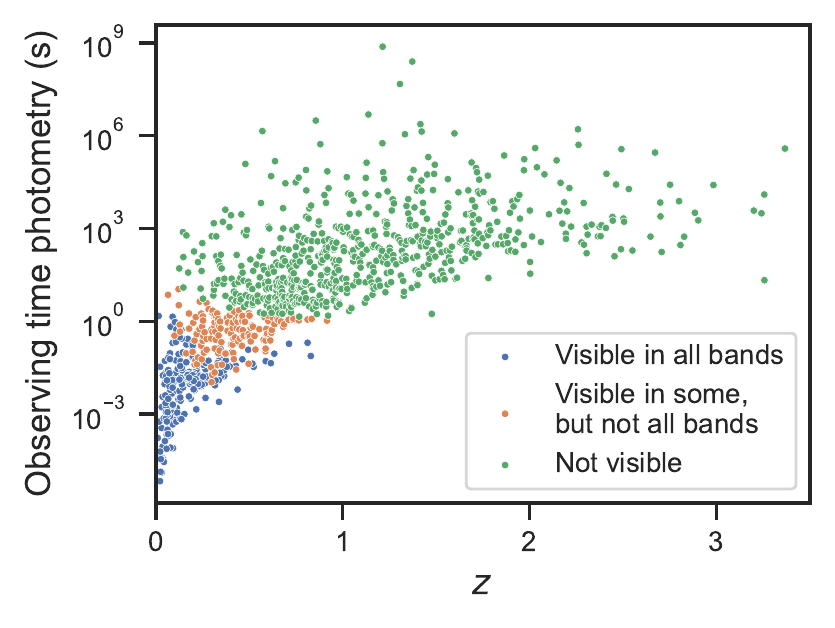}
		\caption{Optical observing time needed for our FRB host galaxies, simulated for CHIME. Photometric observing times are calculated to obtain an SNR of 10 in a single band in a 10-m class telescope, assuming the simulated SDSS r-band magnitudes and a fixed galaxy diameter around \SI{10}{\kilo\pc}. Spectroscopic observing times are larger by a factor 218, which we obtained assuming a slit spectrometer and \SI{1}{\nm} line width (see text for details). Colours represent galaxy visibilities in SDSS. The time varies by 15 orders of magnitude. Blue points have a known photo-$z$. Follow-up strategies should first target orange points. The fixed magnitude limit produced a horizontal cut, therefore the expected observing time for green points is comparable at redshifts where some galaxies are observed (here $z=\numrange[range-phrase = {\text{--}}]{0}{0.7}$).}
		\label{fig:obs_time}
	\end{figure}
	\begin{figure}
		\includegraphics[width=\columnwidth]{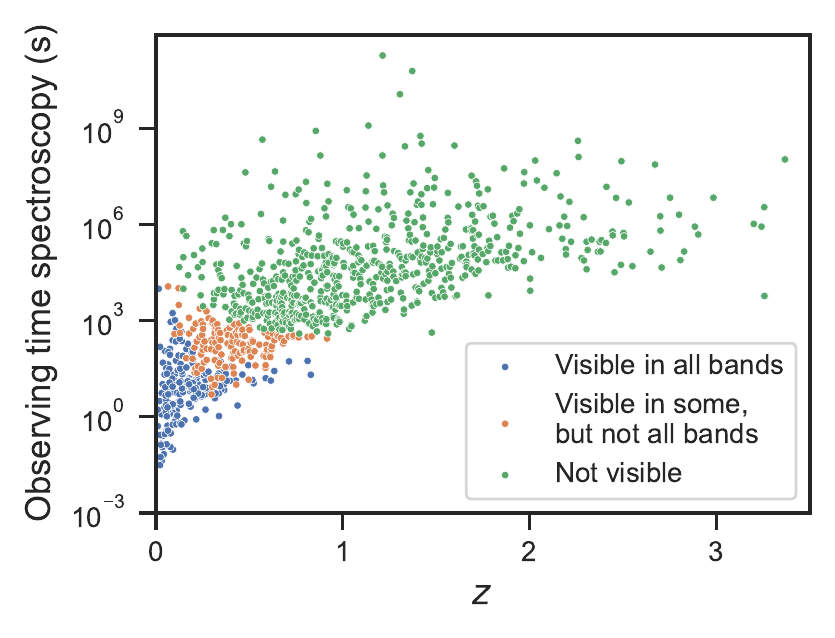}
		\caption{Like Fig.~\ref{fig:obs_time}, but with spectroscopic observing time.}
		\label{fig:obs_time_spec}
	\end{figure}
	\begin{figure}
		\includegraphics[width=\columnwidth]{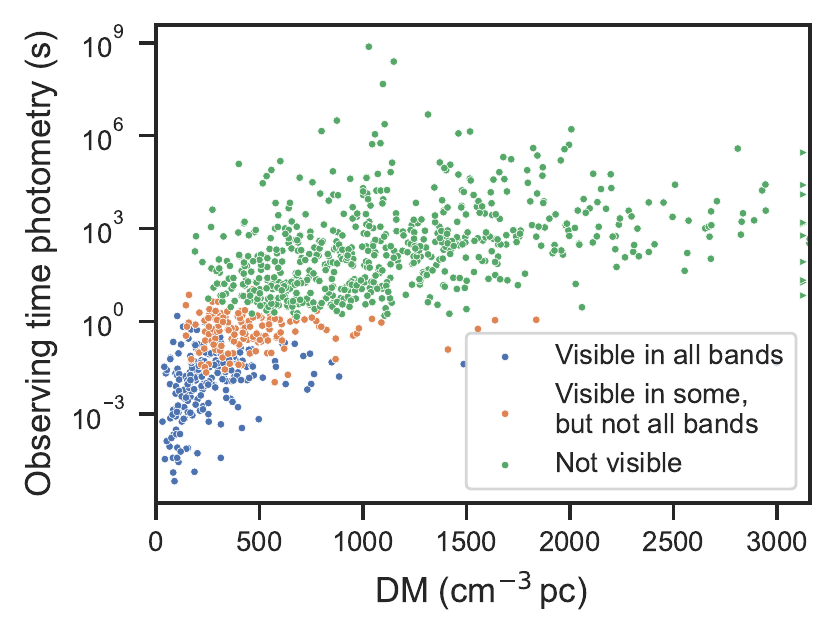}
		\caption{Like Fig.~\ref{fig:obs_time}, but with the \DM{} on the abscissa. The \DM{} will be the best estimate for the distance, as the redshift is the object of desire and therefore unknown a priori. The limits on the \DM{} axis have been chosen, such that it covers the same range as Fig.~\ref{fig:obs_time} in terms of $\langle\DM_\mr{IGM}\rangle(z)$. Triangles indicate points outside the range.}
		\label{fig:obs_time_DM}
	\end{figure}
	
	The photometric observing time is shown in Fig.~\ref{fig:obs_time} against the redshift.
	The observing time required to detect any given galaxy can vary by over \num{15} orders of magnitude, demonstrating that a good observing strategy is necessary.
	The three considered cases -- where FRB host galaxies are not visible, visible in some bands, or visible in all bands -- are clearly separated in different ranges of observing time and redshift.
	
	Galaxies that are visible in SDSS would be visible within seconds in our 10-m example telescope.
	Within a few minutes, one could already make secure associations up to $z\sim 1$.
	The highest expectable observing time is on the order \SI{e3}{\s}.
	As a result, already at $z<1$, a few per cent of FRBs will be without an observable host.
	Above $z\gtrsim 1.5$ a significant fraction will be undetectable by ground based telescopes, false associations would be problematic and make secure associations difficult.
	
	Fig.~\ref{fig:obs_time_spec} shows the observing time needed with our spectrometer.
	Follow-up of this kind would only be possible for bright galaxies that are already visible in at least some bands of SDSS.
	Furthermore, it is limited to $z\lesssim 0.7$.
	
	\subsection{Follow-up optimization for CHIME}
	
	Given the secure detection and lower required observing time of galaxies that are visible in some bands (orange points in Fig.~\ref{fig:obs_time}), it will be most efficient to follow these up first;	at least under the assumption that a high-$z$ population is not required for a given application.
	After these galaxies have been followed up, there is an almost vertical cut below the not visible galaxies (green points), in our example at redshifts 0--0.7.
	This cut results in very similar expected observing times at these redshifts. 
	Generally formulated, the expected time is similar for redshifts where FRB host galaxies have already been found.
	To maximize the cosmological signal, we therefore expect that the most efficient strategy would be to first target the higher redshift host galaxies within this interval, i.e.\ around $z= 0.5$--$0.7$.
	We will test this hypothesis in Section~\ref{sec:optimization}.
	
	In practice, the distance of an FRB is not known a priori but needs to be estimated from FRB properties.
	The DM is already used as a distance estimator on a regular basis \citep{James2022a}.
	The probability density $p(z|\DM)$ has a long tail towards low redshifts, but drops down quickly towards higher redshifts.
	Therefore, the follow-up will often yield host galaxies that are much closer than expected but not much further way.
	The tendency that FRB DMs scatter more towards higher DMs can be seen in Fig.~\ref{fig:obs_time_DM}, where we show the observing time plotted against the DM instead of redshift.
	This asymmetry limits the number of cases where the required follow-up time is much longer than expected.
	Other distance estimators could be the amount of scatter or the width of a burst, but both have a large intrinsic randomness compared to their distance dependence \citep[see e.g.][]{Ocker2022a,CHIME2021}.
	
	As a final note, the large difference in observing time shows that it will not be uncommon to have single galaxies that need significantly more follow-up time than others, or will not be visible at all.
	For example, even at $z\sim 0.3$, there are galaxies who need of order 1000 times longer observing times than even the dimmest galaxies visible in SDSS.
	
	\subsection{Optimal DM and observing time limit}
	\label{sec:optimization}
	
	We want to test the hypothesis that the best target redshift is at the higher end of observed redshifts and investigate which maximum observing time should be spent on one galaxy.
	We further aim to find the best balance between this maximum observing time and the number of FRBs in a sample in the case of limited observing time, and we consider specific follow-up strategies.
	To be independent of the cosmological parameter to be constrained (e.g.\ $\Omega_\mr{m}$ or $H_0$) and for computational feasibility, we change from the MCMC approach to a simpler approach.
	We define the `cosmological signal' of all FRBs with a redshift measurement as
	\begin{equation}
	\mr{SNR}_\mr{c}\equiv\sqrt{\sum_i\left(\frac{\langle\DM_\mr{IGM}\rangle(z_i)}{\sigma_\mr{c}(z_i)}\right)^2}\,,
	\end{equation}
	where $\langle\DM_\mr{IGM}\rangle(z_i)$ is given by Equation~(\ref{eq:Macquart}) and
	\begin{align}
	\sigma_\mr{c}(z_i)&=\sqrt{(\sigma_\mr{h}/(1+z_i))^2+\sigma_\mr{IGM}(z_i)^2}\;\text{, with}\\
	\sigma_\mr{h}&=\DM_0\sqrt{e^{2\sigma_\mr{host}}-e^{\sigma_\mr{host}}}\qquad\text{and}\\
	\sigma_\mr{IGM}(z_i)&=0.2\,\langle\DM_\mr{IGM}\rangle(z_i)/\sqrt{z_i}\,.
	\end{align}
	The estimate for $\sigma_\mr{IGM}$ has been derived from simulations by \citet{Kumar2019} and is valid until $z\sim 3$.
	In this section, we will consider $\mr{SNR}_\mr{c}^2$ because the data will always build on some previous data set with $\mr{SNR}_0$ and therefore yield an improvement
	\begin{equation}
	\frac{\mr{SNR}_\mr{tot}}{\mr{SNR}_0} = \frac{\sqrt{\mr{SNR}_0^2+\mr{SNR}_\mr{c}^2}}{\mr{SNR}_0} \approx 1 + \frac{1}{2}\frac{\mr{SNR}_\mr{c}^2}{\mr{SNR}_0^2} \,,
	\end{equation}
	where $\mr{SNR}_\mr{tot}$ is the total cosmological signal.
	
	We use the FRB host galaxies that are not visible in any of the bands from the previous section (green points in Fig.~\ref{fig:obs_time_DM}).
	To find the best target DM for carrying out the optical follow-up, we pick several DMs and select the 100 FRBs closest to them.
	Some of the galaxies are too dim to be detected in a reasonable time, so an efficient strategy always has to include some upper limit on the observing time that is spent per galaxy.
	Since we do not know what the best time limit would be, we start low and increase the limit gradually until we detect all galaxies.
	For each central DM and time limit, we compute the efficiency $\mr{SNR}_\mr{c}^2/t_\mr{tot}$, where $t_\mr{tot}$ is the total observing time spent on all galaxies.
	
	\begin{figure}
		\includegraphics[width=\columnwidth]{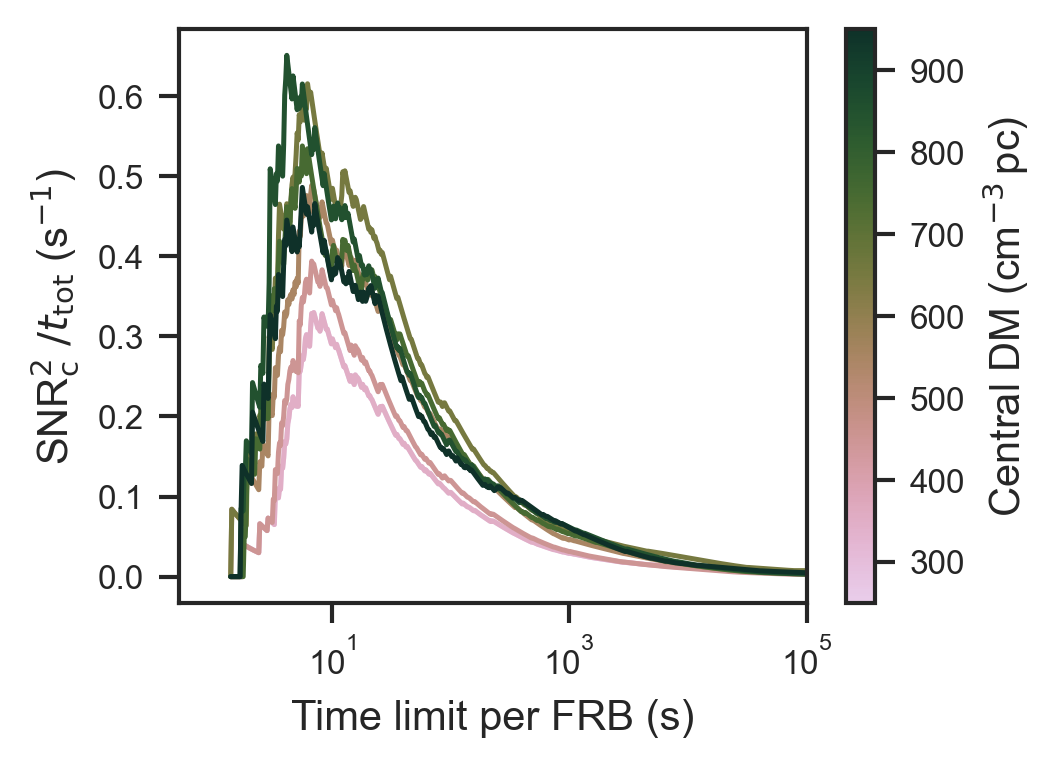}
		\caption{The efficiency as a function of the maximum time spent per host galaxy for different central DMs, each time considering the 100 closest FRBs. The data are the CHIME FRBs, with galaxies not visible in SDSS, i.e. the green dots in Fig.~\ref{fig:obs_time_DM}. }
		\label{fig:efficiency_DM}
	\end{figure}
	This efficiency is shown in Fig.~\ref{fig:efficiency_DM} against the time limit for a few different central DMs.
	The highest efficiency is reached at a low time limit, when only a fraction of the FRBs are observed.
	This can be understood from the distribution in Fig.~\ref{fig:obs_time_DM} remembering that the time axis is logarithmic, indicating a distribution dominated by low observing times with a very long tail.
	To determine the DM centre that can give the highest efficiency, we do smaller DM steps and compute the maximum efficiency for each DM.
	\begin{figure}
		\includegraphics[width=\columnwidth]{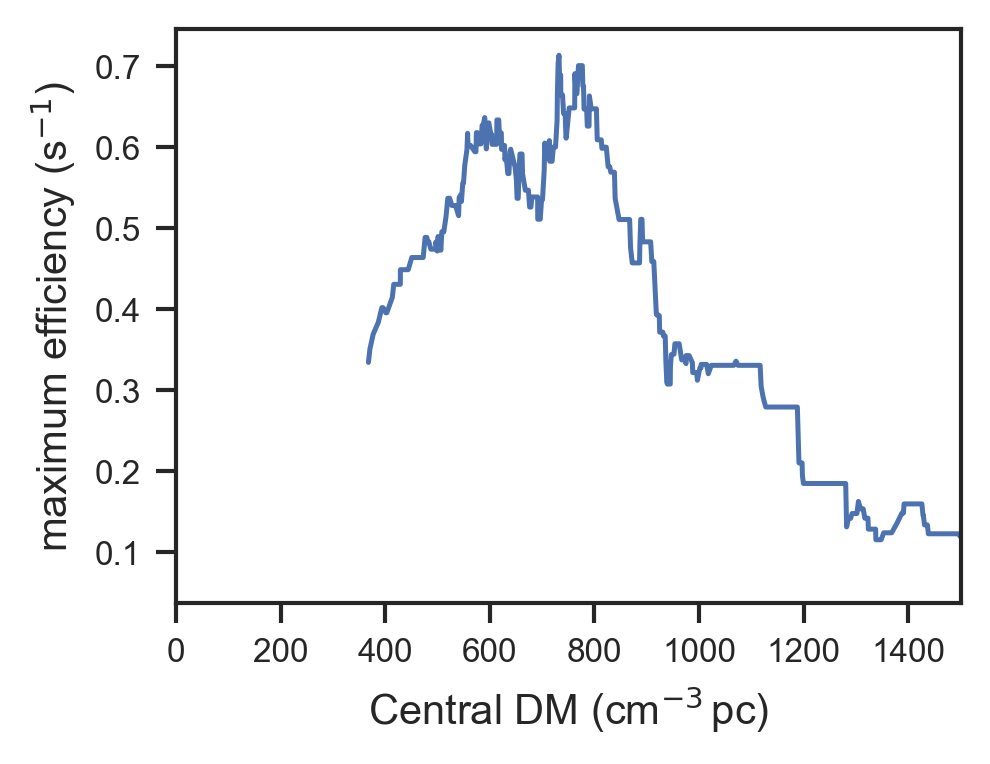}
		\caption{The efficiency at the time limit where it takes on its maximum value. The highest efficiency can be reached around a DM of \SI{750}{\dmu}, but it is relatively constant between 500 and \SI{800}{\dmu}.}
		\label{fig:max_efficieny_DM}
	\end{figure}
	The result is shown in Fig.~\ref{fig:max_efficieny_DM}.
	The highest efficiency is reached around a central DM of \SI{750}{\dmu} in agreement with our predictions in the previous Section, but stochastic variations dominate in the range from \num{\sim500} to \SI{800}{\dmu}.
	
	\begin{figure}
		\includegraphics[width=\columnwidth]{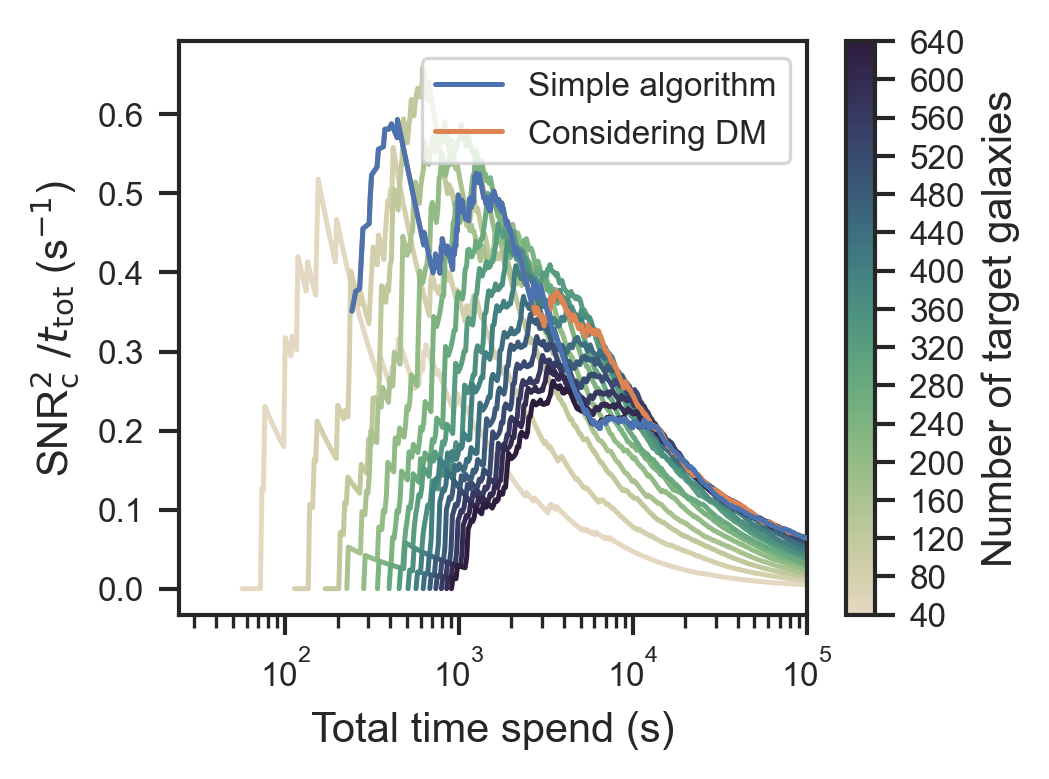}
		\caption{The efficiency in terms of the squared cosmological signal per total observing time. Over-plotted lines show the outcomes of our two developed algorithms. The simple algorithm considers the last four detections, the number of total and detected galaxies, and the total observing time. The order in which galaxies are added is based on the FRB DMs. The second algorithm additionally considers the DM to compute the time limit for FRBs with a DM beyond \SI{1000}{\dmu}.
		}
		\label{fig:opt_time}
	\end{figure}
	When observing time is the limiting factor, we need to balance the FRB sample size that we follow up, against the maximum time spent on each source.
	Since we just found the optimal DM to be around \SI{700}{\dmu}, we consider this finding, but for simplicity only try to maximize the number of detected host galaxies instead of $\mr{SNR}_\mr{c}^2$.
	To consider the previous findings, we start at $\DM=\SI{700}{\dmu}$ and increase our number of FRBs $N$ in the sample gradually by whichever FRB's DM is closest to \SI{700}{\dmu}, but below \SI{1000}{\dmu} until all FRBs below this limit are included.
	We show the efficiency of the detected cosmological signal, $\mr{SNR}_\mr{c}^2/t_\mr{tot}$, against the total time spent, in Fig.~\ref{fig:opt_time} for different $N$.
	For a given observing time, one could read the optimal $N$ from the graph.
	However, if the distribution of galaxies with respect to their required observing time is not known, we propose the following algorithm, which we derive in Appendix~\ref{app:algorithm}.
	
	\begin{itemize}
		\item Start with a number of targets that is small compared to the available observing time but large enough to not be affected by low number statistics, and observe `simultaneously' until the first galaxy is visible.
		\item If the probability $p_{\tl}$ to find a galaxy in the next $\Delta t$ at the current observing time limit $\tl$ satisfies
		\begin{equation}
		p_{\tl}>\frac{n\,(N-n)}{t_\mr{tot}}\,, \label{eq:algo}
		\end{equation}
		where $n$ is the number of detected galaxies, increase $\tl$ until the next galaxy is detected, otherwise increase the sample of target galaxies $N$ by one.
		To estimate $p_{\tl}$, one can take the difference $\Delta \tl$ between the times needed to discover the last $\Delta n$ galaxies and obtain $p_{\tl}=\Delta n/\Delta \tl$.
		\item Repeat this step until the available time runs out.
	\end{itemize}
	In this way, the algorithm essentially finds the optimal $\tl$ and subsequently increases $N$.
	This simple algorithm works well at first, as can be seen in the blue curve in Fig.~\ref{fig:opt_time}.
	However, once the FRBs that are added exceed $\DM\sim\SI{1000}{\dmu}$, the distribution of observing times differs too much towards longer times from the distribution of already observed galaxies.
	Starting at this DM, we impose a second condition, assuming that the distribution roughly keeps its shape.
	\begin{itemize}
		\item If the next FRB is at a $\DM>\SI{1000}{\dmu}$, we require that
		\begin{equation}
		\tl \geq \frac{D_L(\DM)^4}{D_L(\SI{1000}{\dmu})^4}\,t_{\mr{l},700}\,,
		\end{equation}
		where $t_{\mr{l},1000}$ is the time limit when reaching $\DM=\SI{1000}{\dmu}$, and $D_L(\DM)$ is the expected luminosity distance for a given DM, obtained by inverting Equation~(\ref{eq:Macquart}) to get $z(\DM)$.
	\end{itemize}
	The result of the improved algorithm is again shown in Fig.~\ref{fig:opt_time}, yielding close to optimal results at all times.
	Deviations from the ideal efficiency come from our assumption that the highest number will also lead to the highest cosmological signal, but also from the unavoidable fact that the algorithm only knows `past' detections and not the whole population.
	
\begin{figure}
		\includegraphics[width=\columnwidth]{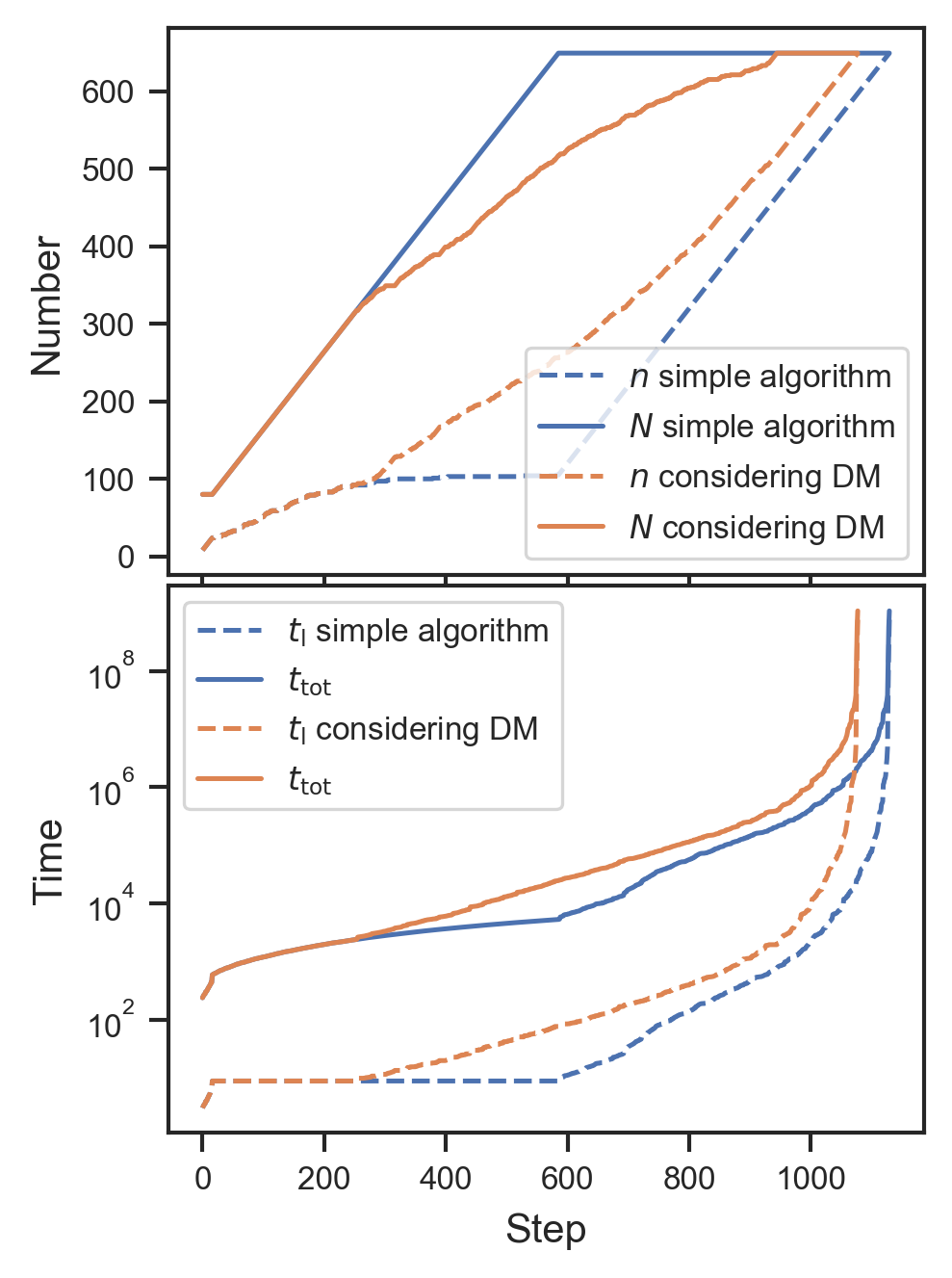}
		\caption{Evolution of the numbers and times in our algorithms. $N$ denotes the FRBs or target galaxy sample size, $n$ the number of detected galaxies, $\tl$ the time limit on each galaxy, and $t_\mr{tot}$ the total observing time spent. In every step, either $\tl$ is increased to detect one more galaxy in the current sample, or one galaxy is added to the sample.}
		\label{fig:algorithm}
	\end{figure}

	The outcomes of the two algorithms are shown in Fig.~\ref{fig:algorithm} in terms of the times and numbers after each step.
	The curve for $\tl$ in the simple algorithm shows that $\tl$ is independent of $N$, as it stays constant with growing $N$ after it is found.
	The theoretical reason is given in Appendix~\ref{app:algorithm}.
	
	\section{Discussion}
	\label{sec:discussion}
	
	\subsection{Limitations}
	Our estimates here are limited by knowledge about the FRB population in several ways.
	The most important uncertainties are the FRB's $L_\mr{max}$ (or $E_\mr{max}$), spectral index, and distance distribution (if it, e.g., follows SFR or SMD).
	The strong dependence on $L_\mr{max}$ is visible in the middle panel of Fig.~\ref{fig:intrinsics}, where a larger fraction of high-$L$ FRBs are observed, compared to less luminous FRBs (this also illustrates why constraints on $L_\mr{max}$ are much better than on $L_\mr{min}$ \citep{James2022}).
	The value of $L_\mr{max}$ directly affects the maximum redshift at which FRBs can be observed.
	The shape of the redshift distribution is rather unaffected by these high-$L$ FRBs, because they are distributed across redshifts.
	
	The spectral index also has a strong influence on the maximum possible observed redshift.
	For example, a burst at $z=2$ observed on Earth between $\nu=1.2$--$\SI{1.4}{\GHz}$ must have been emitted between $\nu=3.6$--$\SI{4.2}{\GHz}$.
	Extrapolating the uncertain value of $\alpha$ to these high frequencies yields a large uncertainty in the maximum observable redshift.
	
	The effect of our different distance models that follow SFR, SMD, and $V_C$ is evident in Figures~\ref{fig:ASKAP}, \ref{fig:CHIME}, and \ref{fig:SKA}, and has already been discussed in Section~\ref{sec:results}.
	
	\subsection{Consequences for FRB applications}
	
	The epochs of H and He \textsc{ii} reionization are expected to be at $z\sim 6$ and $z\sim 3$, respectively.
	FRBs need to be detected from these distances, and their redshift must be obtained.
	The epoch of H reionization cannot be reached by any of our simulated surveys, not even SKA1-Mid.
	While ASKAP's FRBs are also not distant enough to reach the epoch of He \textsc{ii} reionization, CHIME is just reaching it, but only if the cosmic FRB density follows the SFR.
	SKA1-Mid will reach the epoch of He \textsc{ii} reionization in all distance models.
	However, none of the optical surveys detects galaxies at $z > 2$ in all bands, making dedicated optical follow-up a necessity to detect the He \textsc{ii} epoch of reionization.
	Moreover, we showed in Section~\ref{sec:obstimes} that this optical follow-up is not feasible with a 10-m ground-based telescope, but likely needs to be carried out from space.
	
	In Section~\ref{sec:missing}, we examined the effects of limited follow-up on cosmological constraints, in particular on $\Omega_\mr bh_{70}$.
	When only optical surveys are used to obtain FRB redshifts, the usable FRB population is restricted to low redshifts.
	These low-$z$ FRBs result in lower constraints on $\Omega_\mr bh_{70}$ or in correspondingly more FRBs needed to reach the same constraints.
	Note, that it is not beneficial to increase the number of low-$z$ FRBs indefinitely, as at some number, several FRBs will probe the same sight lines \citep{Reischke2023}.
	
	To probe the intergalactic magnetic fields \citep{Akahori2016}, FRBs have to be distant enough that the intergalactic contribution to the rotation measure becomes comparable to the host contribution.
	Depending on the progenitor of FRBs, this will likely only be the case around $z\gtrsim 3$ \citep{Hackstein2019}.
	SKA1-Mid FRBs with optical follow-up of space based telescopes might therefore be needed to reach sufficient numbers for this method.
	
	The signal from a hypothetical photon mass almost plateaus around $z\sim 1$ \citep[see e.g.\ fig.~1 of][]{WeiWu2020}.
	Therefore, FRBs at $z\lesssim 1$ are best for this application.
	They can be obtained with any of the radio surveys, but need an optical follow-up that is deeper than SDSS.
	
	FRBs that are gravitationally lensed by an intervening galaxy or galaxy cluster are so rare and valuable \citep[see e.g.][]{Wucknitz2021} that they should be followed up in any possible way.
	
	\subsection{Studying FRB progenitors}
	
	Optical follow-up of host galaxies is not only important for FRB applications but also for studies of FRB origins \citep[e.g.][]{Heintz2020,Bhandari2022}.
	Photometric studies can localize FRBs within galaxies, for example to spiral arms or star-forming regions \citep[see e.g.][]{Tendulkar2021}, and allow comparing the morphological types with other transients.
	Spectroscopy can reveal the star formation history via stellar population synthesis.
	These methods mainly require close by FRBs to obtain a uniformly well-studied set of host galaxies and direct environments.
	To obtain a set that is as unbiased as possible, FRB follow-up should be deep with conservative upper limits on DM to not exclude close by high $\DM_\mr{host}$ FRBs.
	
	FRBs at $z>2$ are interesting to study the evolution with the cosmic SFR and the frequency dependence of the rate.
	Our simulations show how the possibility of obtaining a complete set depends on the depth of the optical survey.
	DELVE is already nearly complete at $z\lesssim 0.4$.
	Dedicated follow-up from ground base telescopes could yield nearly complete sets up to $z\sim 1$, depending on the available time.
	Larger redshifts might only be accessible with space based telescopes.
	
	\subsection{Outlook}
	
	In our models, we made a few assumptions that would lead to biases in inferred parameters.
	For example, the expected $\DM_\mr{host}$ will likely correlate with the host galaxy's mass and SFR.
	In turn, brighter galaxies will be biased towards higher $\DM_\mr{host}$, which is not a problem if the properties of the $\DM_\mr{host}$ distribution are inferred together with e.g.\ the missing baryon density.
	However, the bias has to be taken into account when combining FRBs that have been followed up to different depths or with otherwise different strategies.
	Other biases can come from misidentified galaxies.
	The influence of these effects on observed galaxy properties were previously inspected by \citet{Seebeck2021}.
	This study will serve as a basis for the community to investigate biases on FRB applications in the future, but must be complemented by magnetohydrodynamic simulations.
	
	The optimal follow-up time for any given FRB is also affected by these considerations.
	Deeper follow-up will decrease the number of misidentified host galaxies, as the true host might emerge out of the noise.
	Additionally, it will increase the number of identified galaxies close to the line of sight whose halo is intersected by an FRB \citep[see e.g.][]{Simha2020}.
	For the design of an FRB follow-up campaign, these effects need to be considered, to essentially weigh the quantity against the quality of localized FRBs.
	
	A second use-case of the model is getting prior probabilities for host galaxies to be observable.
	The probability of an FRB-host association depends on the prior probability that the true host is below the detection threshold.
	This prior probability could be calculated from our simulations for given radio and optical telescopes and an FRB's DM.
	
	\section{Conclusions}
	\label{sec:conclusion}
	
	How limiting is optical follow-up for FRB applications?
	To answer this question, we have simulated a realistic FRB population, using parameters obtained by recent studies \citep{James2022, Shin2022}.
	We used galaxies from a semi-analytic model as the mock hosts of our FRBs and tested how many would be visible in current and future optical and infrared surveys.
	As representative radio telescopes, we used ASKAP, CHIME, and SKA1-Mid.
	As host galaxy surveys, we used SDSS, DELVE, Euclid, and LSST.
	
	\begin{itemize}
		\item We found that all applications that require FRBs with measured redshifts can be severely limited by the number of detected host galaxies, since e.g.\ only 20--40 per cent of CHIME FRBs within the SDSS footprint are also visible in all of its bands, additionally they are limited to $z<0.5$.
		On the other hand, a deeper survey like DELVE will detect \numrange{63}{85} per cent of ASKAP's FRBs. Although, a detection does not guarantee a secure association.
		
		\item The redshift ranges resulting from our simulation suggest that the He \textsc{ii} epoch of reionization, expected at $z\sim 3$, will be measurable by several radio telescopes.
		However, dedicated space based follow-up will be needed to obtain redshifts, as even LSST is not deep enough to detect most FRB host galaxies at $z\gtrsim1.5$.
		The same restrictions apply to the use of FRBs as a probe of intergalactic magnetic fields, which also requires FRBs at $z\gtrsim 3$.
		The H epoch of reionization around $z\sim6$ can not yet be reached, even with SKA1-Mid.
		
		\item Applying existing methods to constrain the missing baryons, we showed that even if just 524 of 1000 FRB hosts have measured redshifts, $\Omega_\mr b h_{70}$ can be constrained to 10 per cent (with 95 per cent credibility).
		This would be a great improvement over the constraints of 60 percent from current \textsc{O vii} absorption line studies in X-rays \citep{Kovacs2019}, and one step closer to the uncertainties of theoretical predictions of \num{2.3} per cent from big bang nucleosynthesis and 1.3 per cent from big bang nucleosynthesis combined with \citet{Planck2020} cosmic microwave background measurements \citep{Pitrou2018, Driver2021}.
		
		\item Assuming an optical 10-m class telescope and sufficient FRB localization precision, we showed that follow-up with ground based telescopes can only yield secure associations at $z\lesssim 1.5$ and spectra of galaxies at $z\lesssim 0.7$
		
		\item In general, to minimize observing time, the first FRBs to be followed up, should be those whose hosts can be identified in some optical bands of the large surveys.
		Afterwards, galaxies at the higher redshifts at which host galaxies were observed in the optical survey yield the largest cosmological signal per observing time; their shorter required observing time outweighs the larger cosmological signal of high redshift FRBs.
		DMs of FRBs are well suited as a distance estimate for targeting the optimal redshifts.
		Although, resulting biases have to be taken into account.
		We show that the optimal observing time limit is independent of available time or number of FRBs.
		However, it increases when observing galaxies at higher distances than the ones of galaxies visible in optical surveys.
		We provide methods to find the optimal observing time limit.
	\end{itemize}
	
	\section*{Acknowledgements}
	
	We thank Stefan Hackstein for helping shape the project in the beginning, Luiz F. S. Rodrigues for early discussions on \textsc{galform}, Nataliya Porayko for tips on how to test the MCMC simulations, Lachlan Marnoch and Sunil Simha for their input regarding optical surveys, Clancy James for providing details about the ASKAP/CRACO update, David Gardenier as well as \citet{Macquart2020} for making their code available, Olaf Wucknitz for useful comments on the manuscript, and the referee for several useful comments, in particular for suggesting to assume a 10-m telescope instead of arbitrary units.
	
	LGS is a Lise Meitner Max Planck independent research group leader and acknowledges funding from the Max Planck Society. 
	
	CMB acknowledges support from the Science Technology Facilites Council (STFC) through ST/T000244/1. This work used the DiRAC@Durham facility managed by the Institute for Computational Cosmology on behalf of the STFC DiRAC HPC Facility (www.dirac.ac.uk). The equipment was funded by BEIS capital funding via STFC capital grants ST/P002293/1, ST/R002371/1 and ST/S002502/1, Durham University and STFC operations grant ST/R000832/1. DiRAC is part of the UK's National e-Infrastructure.
	\section*{Data Availability}
	
	All code used for this paper is available as a \textsc{python} package, \textsc{mockFRBhosts}, at \url{https://github.com/JoschaJ/mockFRBhosts}. The host galaxies simulated with \textsc{galform} are available at \url{https://zenodo.org/record/7926078}.
	
	
	
	\bibliographystyle{mnras}
	\bibliography{mock_observations} 

	
	
	
	\appendix
	
	\section{Figures of all telescope combinations}
	\label{app:results}
	
	\begin{figure*}
		\includegraphics[width=\textwidth]{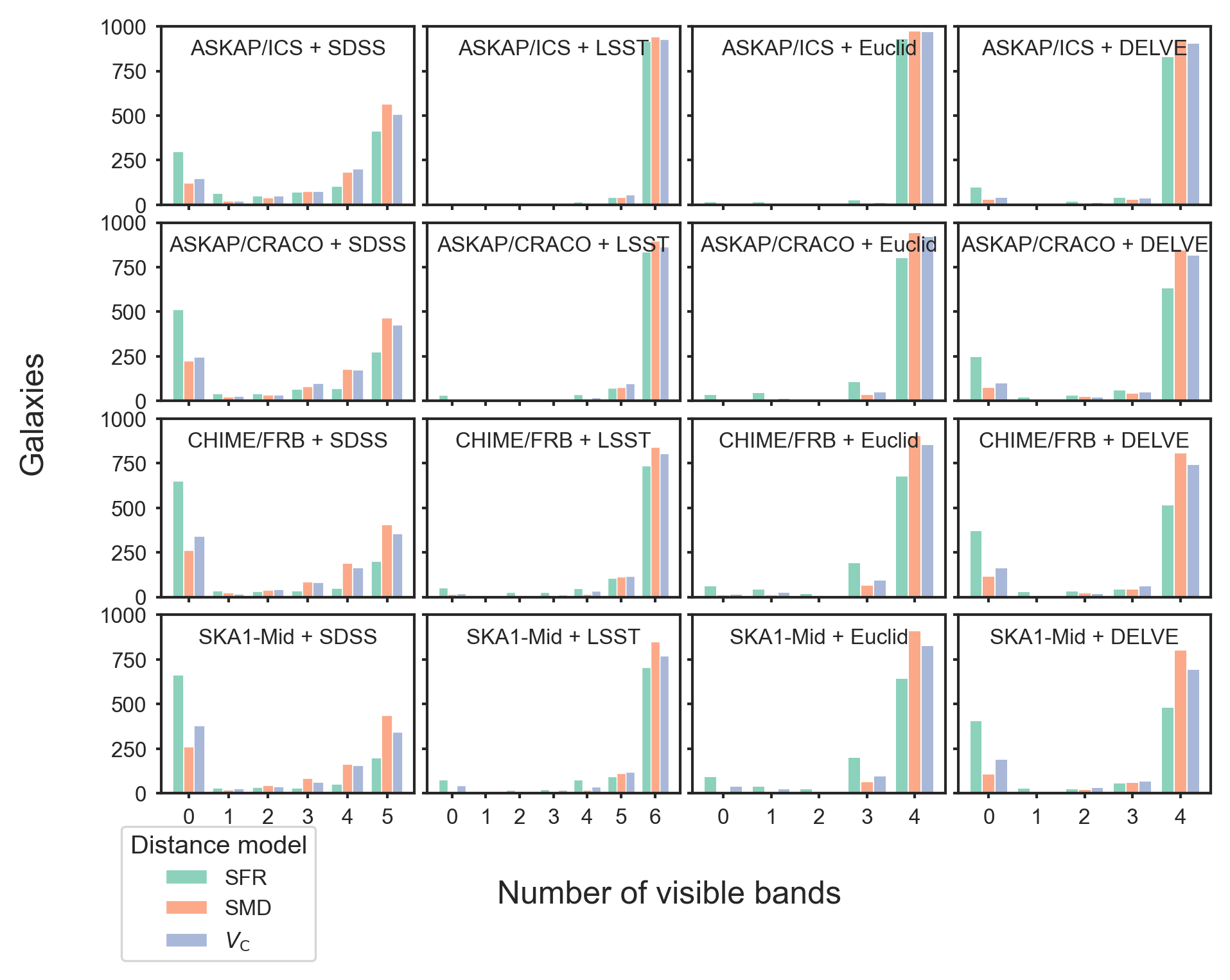}
		\caption{Forecast of the number of pass bands in which FRB host galaxies will be observed for all combinations of radio surveys and optical/infrared surveys that we simulated. Simulations were carried out for three different intrinsic FRB distance distributions, each simulated with \num{1000} FRBs.}
		\label{fig:n_bands}
	\end{figure*}
	
	\begin{figure*}
		\includegraphics[width=\textwidth]{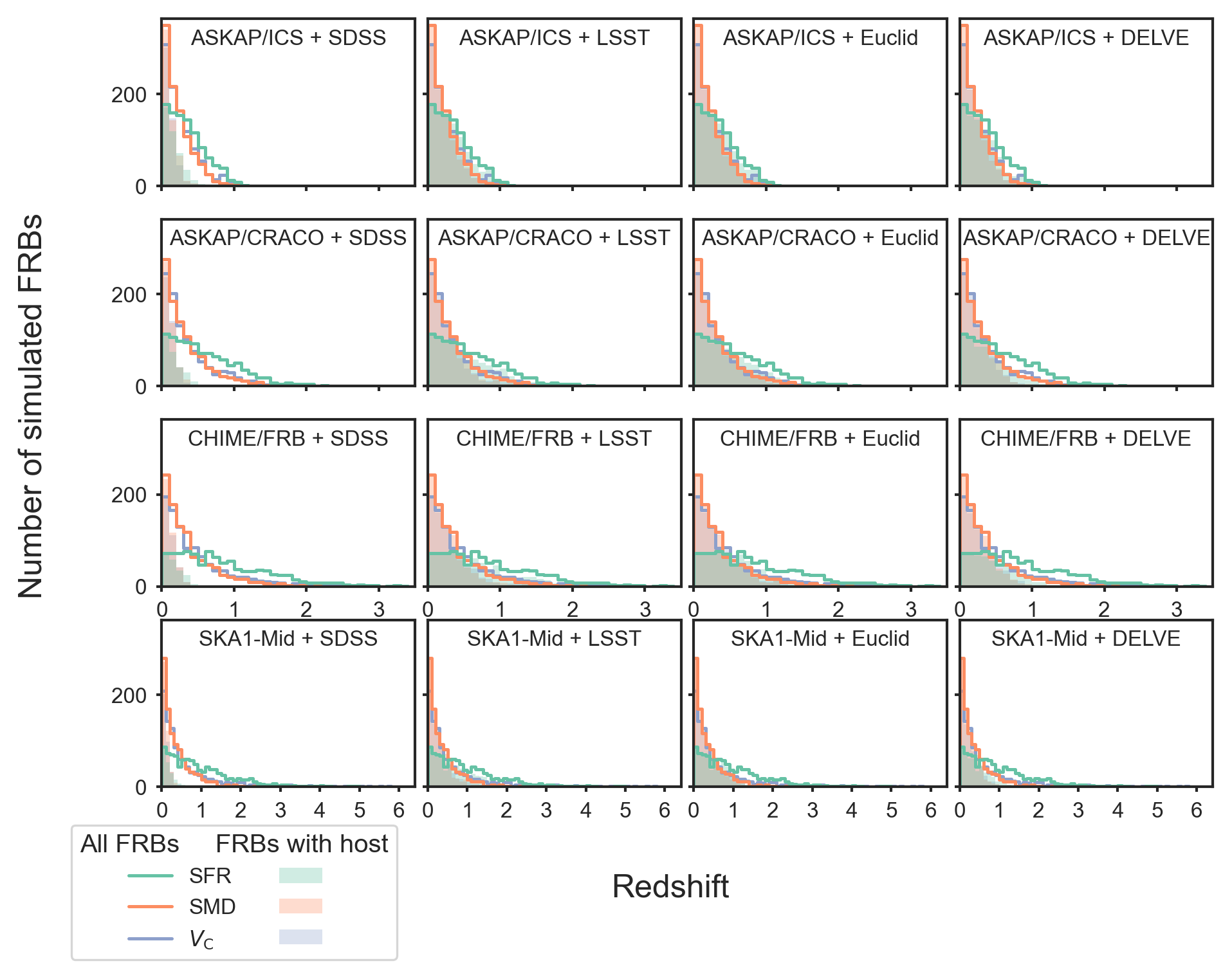}
		\caption{The forecasted redshift distributions of detected FRBs (lines) and of FRBs whose host galaxy was detected in all bands (shaded regions) for all simulated combinations of radio and optical/infrared surveys.}
		\label{fig:z_dists}
	\end{figure*}
	
	\section{Derivation of the optimal time limit}
	\label{app:algorithm}
	To derive Equation~(\ref{eq:algo}), we compare the efficiency of increasing the observation time limit $\tl$ with the efficiency of increasing the galaxy sample $N$. We have to assume that all follow-up times are drawn from the same probability density function, i.e.\ we ignore here the dependence on the DM. The gain in detections from increasing $\tl$ by a small $\Delta \tl$ is given by $\Delta n=p_{\tl}\,\Delta \tl$, where $p_{\tl}$ is just the probability at $\tl$ to find a galaxy in the next $\Delta\tl$. Simultaneously, the total observing time increases by $\Delta t_\mr{tot}\approx(N-n)\,\Delta\tl$, yielding the efficiency
	\begin{equation}
	\left(\frac{\Delta n}{\Delta t_\mr{tot}}\right)_{\tl}=\frac{p_{\tl}}{N-n}\,.
	\end{equation}
	On the other hand the gain from increasing the total number is approximately $\Delta n=n/N\,\Delta N$, and the additional time is $\Delta t_\mr{tot}\approx t_\mr{tot}/N\,\Delta N$, which yields
	\begin{equation}
	\left(\frac{\Delta n}{\Delta t_\mr{tot}}\right)_{N}=\frac{n/N}{t_\mr{tot}/N}=\frac{n}{t_\mr{tot}}\,.
	\end{equation}
	Setting the two equations equal yields the point where increasing $\tl$ is just as efficient as increasing $N$,
	\begin{equation}
	\frac{p_{\tl}}{N-n}=\frac{n}{t_\mr{tot}}\,.
	\end{equation}
	This gives Equation~(\ref{eq:algo}).
	
	In this Equation, the optimal $\tl$ is independent of $N$. To show this, we rearrange the terms and rewrite it as
	\begin{equation}
	\frac{p_{\tl}}{N}\frac{t_\mr{tot}}{N}=\frac{n}{N}\frac{N-n}{N}\,.\label{eq:indep}
	\end{equation}
	For any given $\tl$, each of the fractions is independent of $N$.
	
	If one wants to compute $\tl$ without applying our algorithm, good knowledge of the distribution of observing times is needed. Expressed in terms of the probability density $p(t)$ of finding a galaxy in the observing time interval $\dd{t}$, the expected observed number $n$ after $\tl$ will be
	\begin{equation}
	n=N\int_{0}^{\tl}p(t)\dd{t}\,,\label{eq:n}
	\end{equation}
	and the observing time will be
	\begin{equation}
	t_\mr{tot}=N\int_{0}^{\tl}p(t)t\dd{t}+N\tl\int_{\tl}^{\infty}p(t)\dd{t}=N\tl+N\int_{0}^{\tl}p(t)(t-\tl)\dd{t}
	\label{eq:ttot}
	\end{equation}
	Inserting Equations~(\ref{eq:n}) and (\ref{eq:ttot}) into Equation~(\ref{eq:indep}) and using $p(\tl)=p_{\tl}/N$ (or alternatively taking the derivative of $n/t_\mr{tot}$ with respect to $\tl$) we obtain
	\begin{equation}
	p(\tl)\left(\tl+\int_{0}^{\tl}p(t)(t-\tl)\dd{t}\right) = \int_{0}^{\tl}p(t)\dd{t}\left(1-\int_{0}^{\tl}p(t)\dd{t}\right)\,.
	\end{equation}
	This Equation can be inverted numerically to obtain the optimal $\tl$. Subsequently, one can calculate the optimal $N$ for a given observing time $t_\mr{tot}$ from Equation~(\ref{eq:ttot}).
	
	We can generalize this result to maximize $\mr{SNR}_\mr{c}^2$ instead of the number and further include the DM dependency. 
	The SNR and time will be given in terms of the expected $\mr{SNR}(\DM_\mr i)$ at a given DM, where indices go over all FRBs, by
	\begin{align}
	\mr{SNR}_\mr{c}^2 &= \sum_{i=1}^{N} \int_{0}^{\tl}p(t,\DM_\mr i)\, \mr{SNR}(\DM_\mr i)^2 \dd{t}\,,\quad\text{and}\\
	t_\mr{tot} &= \sum_{i=1}^{N} \left[\int_{0}^{\tl} p(t,\DM_\mr i)t\dd{t} + \tl\left(1-\int_{0}^{\tl} p(t,\DM_\mr i)\dd{t}\right)\right]
	\end{align}
	The maximum condition $\dv{\tl}\frac{\mr{SNR}_\mr{c}^2}{t_\mr{tot}}=0$ yields
	\begin{equation}
	t_\mr{tot}\sum_{i=1}^{N}p(\tl,\DM_\mr i)\, \mr{SNR}(\DM_\mr i)^2 = \mr{SNR}_\mr{c}^2 \sum_{i=1}^{N} \left(1-\int_{0}^{\tl} p(t,\DM_\mr i)\dd{t}\right).
	\end{equation}

	\bsp	
	\label{lastpage}
\end{document}